\documentclass[letterpaper, 10 pt, conference]{ieeeconf}
\usepackage{amsmath,amssymb,amsfonts,dsfont}
\usepackage{tabularx}
\usepackage{cite}
\usepackage{mathtools}
\usepackage{algorithmic}
\usepackage{graphicx}
\usepackage{algorithm,algorithmic}
\usepackage{hyperref}
\hypersetup{hidelinks=true}
\newtheorem{lemma}{Lemma}

\newtheorem{remark}{Remark}

 \newtheorem{theorem}{Theorem}[section]
 
 \newtheorem{corollary}{Corollary}
\newtheorem{assumption}[theorem]{Assumption}

\DeclareMathOperator*{\argmin}{arg\,min}
\DeclareMathOperator{\atantwo}{atan2}

\usepackage{comment}
\usepackage{enumerate}
\IEEEoverridecommandlockouts 

\begin{document}
\title{Shortest Trajectory of a Dubins Vehicle with a Controllable Laser}
\author{Shivam Bajaj$^{1}$, Bhargav Jha$^{2}$, Shaunak D. Bopardikar$^{2}$, Alexander Von Moll$^{3}$, and David W. Casbeer$^{3}$
\thanks{This research was supported by the Aerospace Systems Technology Research and Assessment (ASTRA) Aerospace Technology Development and Testing (ATDT) program at AFRL under contract number FA865021D2602.
Distribution Statement A: Approved for public release.  Distribution is unlimited.  Case numbers AFRL-2023-4780 and 2023-0941. }
\thanks{S. Bajaj is with the Electrical and Computer Engineering Department, Purdue University, West Lafayette, IN 47906 USA (e-mail: bajaj41@purdue.edu).}
\thanks{B. Jha, and S.D. Bopardikar are with the Electrical and Computer Engineering Department, Michigan State University, East Lansing, MI 48823 USA (e-mail: bhargav@msu.edu, shaunak@egr.msu.edu).}
\thanks{A. Von Moll and D.W. Casbeer are with Control Science
Center, Air Force Research Laboratory, Ohio, USA (e-mail: alexander.von\_moll@afrl.af.mil, david.casbeer@us.af.mil).}}

\maketitle

\begin{abstract}
We formulate a novel planar motion planning problem for a \emph{Dubins-Laser} system that consists of a Dubins vehicle with an attached controllable laser. The vehicle moves with unit speed and the laser, having a finite range, can rotate in a clockwise or anti-clockwise direction with a bounded angular rate. From an arbitrary initial position and orientation, the objective is to steer the system so that a given static target is within the range of the laser and the laser is oriented at it in minimum time. We characterize multiple properties of the optimal trajectory and establish that the optimal trajectory for the Dubins-laser system is one out of a total of 16 candidates. Finally, we provide numerical insights that illustrate the properties characterized in this work.
\end{abstract}


\section{Introduction}\label{sec:introduction}
Over the last few years, there has been an accelerated demand for Unmanned Aerial Vehicles (UAVs) due to several applications such as in agriculture, search and rescue, monitoring, and security \cite{shakhatreh2019unmanned,kim2019unmanned,hayat2017multi,9993007}. It is now well known that using cooperative heterogeneous agents, for instance UAVs and ground robots, in such applications improves the overall performance \cite{parker2016exploiting,mukhamediev2023coverage}. Generally, in these applications, the motion of one agent is independent of the motion of the other agent. However, some applications may require a careful coordination between the agents. For instance, a camera attached to a UAV, which can be viewed as a combination of two heterogeneous agents, is used for applications such as aerial photography and surveillance \cite{rajesh2015camera}. In security applications, a similar setup can be considered with a laser or a turret\cite{bajaj2023perimeter,von2021turret} attached to a UAV which motivates further research in joint motion planning of heterogeneous agents.

In this work, we formulate a novel joint motion planning problem for a UAV with an attached laser. In particular, the UAV is modelled as a Dubins vehicle, i.e., a nonholonomic vehicle that is constrained to move along paths of bounded curvature and cannot reverse its direction~\cite{dubins1957curves}. The laser has a finite range and can rotate clockwise and anti-clockwise while being attached to the vehicle. Henceforth, we will denote the Dubins vehicle with an attached controllable laser as the Dubins-Laser (Dub-L) system. The aim of this work is to determine a time optimal trajectory for the Dub-L system such that a static target is within the range of the laser and the laser is oriented towards the target in minimum time. 

The problem of finding the shortest path between two specified locations and orientations at those locations for a Dubins vehicle was first characterized geometrically by Dubins in \cite{dubins1957curves}. A landmark result from \cite{dubins1957curves} states that the optimal path for such a vehicle is a concatenated segment of circular arcs $C$ and straight line segments $S$. Specifically, the shortest path is of type $CSC$ or $CCC$ or a subsegment (for instance $CS$ or $CC$) of these two types of paths.
The concatenating points on the trajectory where the Dubins vehicle switches from one type of segment to another type of segment are commonly known as the \emph{switching points}.
If the orientation at the final location is not specified, then the problem is generally known as the \emph{relaxed Dubins problem} and the shortest path is of type $CS$ or $CC$ or a subsegment of these two \cite{bui1994accessibility}.
As an extension of the Dubins vehicle, \cite{reeds1990optimal} considered a vehicle which goes forward as well as backwards. Later, \cite{bui1994shortest} and \cite{sussmann1991shortest} independently derived the same result as in \cite{reeds1990optimal} by using optimal control theory. In both these works, it is established that the optimal control problem is \emph{normal}, i.e., the costate multiplied with the cost function is non-zero. Given that the Dubins vehicle is equivalent to the vehicle considered in \cite{reeds1990optimal}, with an exception that the vehicle in \cite{reeds1990optimal} goes backwards, 
the authors in \cite{bui1994shortest} then assume that the optimal control problem for the Dubins vehicle is also normal and thus, establish the same result as in \cite{dubins1957curves} via optimal control theory. This implicit assumption has also been made in several other works \cite{bui1994accessibility,ma2006receding,jha2022time}.
Recently, \cite{kaya2017markov} established that the optimal control problem considered in \cite{dubins1957curves} is \emph{abnormal}, i.e., the costate multiplied with the cost function may be zero. Further, \cite{kaya2017markov} also characterized the \emph{abnormal solution} for the Dubins vehicle, i.e., the set of paths obtained when the costate associated with the cost function is zero.

Numerous variations \cite{bakolas2013optimal,bui1994accessibility,isaiah2015motion,chitsaz2007time,ny2011dubins} of the optimal control problem studied in \cite{bui1994shortest} have been extensively studied including obstacle avoidance \cite{boissonnat1996polynomial,jones2021generalization,jha2022time}, path planning \cite{lin2014path}, intercepting targets \cite{liu2014evasion, gupta2022shortest}, target tracking \cite{anderson2014stochastic,kokolakis2023bounded}, and coverage problems \cite{karapetyan2018multi}. 
Many robotic path planning applications, especially coverage problems, require a camera attached to a UAV \cite{jakob2010occlusion,kim2010uav,tzes2018visual,schwager2011eyes,schwager2009optimal}. These works assume that the orientation of the camera is fixed, generally in the downward direction. However, a camera that has the ability to rotate while being attached to the UAV may improve the performance, such as reducing the total time to reach a location. Recently, \cite{sabetghadam2019optimal,papaioannou2022integrated,bousias2019collaborative} considered a gimbaled camera with ability to pan and tilt while being attached to the UAV, which is modelled either as a single or a double integrator. The main contributions of this work are:

\begin{enumerate} 
    \item \textbf{Optimal Control of Dubins-Laser system:} We formulate a novel optimal control problem of a Dubins vehicle with an attached laser system. The laser, having a finite range, is modelled as a single integrator and rotates either clockwise or anti-clockwise in the environment. The environment consists of a single static target. The aim is to determine an optimal control strategy of the Dub-L system such that the target is within the range of the laser and the laser is oriented towards the target in minimum time.
    
    \item \textbf{Set of Candidate Paths:} We establish that the number of candidate trajectories can be at most 16. 
    \item \textbf{Semi-Analytical Solution of the Optimal Path:} Finally, we establish that the optimal trajectory can be efficiently determined by solving a set of nonlinear equations, hence providing a semi-analytical solution to the problem.
\end{enumerate}

This work is organized as follows. In Section \ref{sec:prob}, we formally pose the problem and in Section \ref{sec:necessary}, we apply the maximum principle to characterize the optimal control for the Dub-L system. Section \ref{sec:charac_shortest_path} characterizes multiple properties of the shortest trajectory and establishes the main result of this work. In Section \ref{sec:sol}, we determine the semi-analytical solution of the shortest trajectory and Section \ref{sec:numerics} provides multiple numerical simulations to illustrate the properties established in this work. Finally, Section \ref{sec:conclusion} summarizes this work and outlines possible future directions.

\section{Problem Description}\label{sec:prob} 
\begin{figure}
    \centering
\includegraphics[scale=1.8]{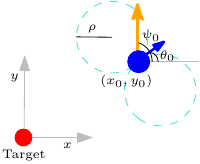}
    \caption{Problem Description. The red dot depicts the static target located at the origin and the blue dot represents the Dubins vehicle. The laser is depicted by the yellow (longer) arrow and the blue (shorter) arrow depicts the orientation of the Dubins vehicle.}
    \label{fig:Problem_desc_dubins}
\end{figure}

We consider the problem of optimally steering a Dub-L system in the plane that consists of a Dubins vehicle and a controllable laser which is attached to the vehicle (refer to Fig. \ref{fig:Problem_desc_dubins}). The laser, modelled as a single integrator, has a finite range $r>0$. The vehicle, having a minimum turn radius $\rho>0$, moves with a constant unit speed and the laser has the ability to rotate either clockwise or anti-clockwise with speed $\omega\leq \omega_M$, for a given $\omega_M>0$.
The environment consists of a static target, assumed to be located at the origin $(0,0)$. The state vector $\mathbf{x}\in \mathbb{R}\times\mathbb{R}\times\mathbb{S}^1\times\mathbb{S}^1$ of the Dub-L system is $\textbf{x} = \begin{bmatrix}x & y & \theta & \psi\end{bmatrix}^\top$, where $(x,y)\in \mathbb{R}^2, \theta \in \mathbb{S}^1$, and $\psi \in \mathbb{S}^1$ denotes the location of the vehicle, the heading direction of the vehicle measured in the anticlockwise direction from the positive $X$-axis, and the orientation of the laser measured in the anticlockwise direction from the positive $X$-axis, respectively. The kinematic equations that describe the motion of the Dub-L system are
\begin{align*}
    \dot{x} = \cos(\theta),\ \dot{y} = \sin(\theta),\ \dot{\theta} = \frac{u}{\rho},\ \dot{\psi} = \frac{u}{\rho}+\omega
\end{align*}
where, at any given time instant $t$, $u(t)\in [-1,1]$ denotes the control input of the Dubins vehicle and $\omega(t) \in [-\omega_M,\omega_M]$ denotes the angular speed of the laser. We denote the control vector as $\textbf{u}=\begin{bmatrix}
    u & \omega
\end{bmatrix}^\top$ and is defined as $\textbf{u}:[0, \infty)\to \mathbb{R}^2$. 

The aim of the Dub-L system is to \emph{capture} the target in minimum time, where the target is said to be captured when the distance between the target and the Dub-L system is at most $r$ and the laser is oriented towards the target. Formally, the target is said to be captured at final time $t_f$ if the following two conditions jointly hold:
\begin{subequations}\label{eq:terminal_constraints}
\begin{equation}
\label{eq:terminal_constraints_a}
x^2(t_f)+y^2(t_f)\leq r^2
\end{equation}
\begin{equation}
\label{eq:terminal_constraints_b}
\pi+\atantwo\left(y(t_f),x(t_f)\right) -\psi(t_f) = 0
\end{equation}
\end{subequations}
where $\atantwo(\cdot,\cdot):\mathbb{R}\times \mathbb{R}\rightarrow [0,2\pi]$ is the four quadrant inverse tangent function. In this work, we assume that the initial location $(x(0),y(0))$ of the Dub-L system is such that $x^2(0)+y^2(0)>r^2$ holds at the initial time.

The trajectory of the Dub-L system is comprised of the trajectory of the Dubins vehicle (corresponding to states $x$, $y$, and $\theta$) as well as the trajectory of the laser (corresponding to the state $\psi$). We refer to the 4-tuple $\left(x(t),y(t),\theta(t),\psi(t)\right)$ as \textit{trajectory} of the Dub-L system, the 3-tuple $\left(x(t),y(t),\theta(t)\right)$  as the \emph{pose trajectory} of the Dub-L system, and $\psi(t)$ as \emph{laser's trajectory}. 
To succinctly denote the trajectory of the Dub-L system, we will use the notation $\mathcal{T}(t) = \Gamma(t)|\mathcal{L}(t)$, where $\Gamma(t)$ (resp. $\mathcal{L}(t)$) denotes the pose (resp. laser's) trajectory. For instance, if the pose trajectory is a straight line segment $S$ and the laser's trajectory is to rotate clockwise, then $\mathcal{T}(t) = S|+$, where $+$ (resp. $-$) denotes that the laser turns clockwise (resp. anti-clockwise)\footnote{In Section \ref{sec:necessary}, we will see that the pose trajectory of the Dub-L system comprises circular arcs and straight line segments.}.

Given an initial state $\textbf{x}(0):=\textbf{x}_0=\begin{bmatrix}x_0 &y_0 &\theta_0&\psi_0\end{bmatrix}^{\top}$ of the Dub-L system,  the objective of this work is to determine a time-optimal trajectory for the Dub-L system such that equations \eqref{eq:terminal_constraints} hold for the least possible final time $t_f$. 

The proposed optimal control problem may have non-unique solutions for some initial conditions and problem parameters. For instance, let $\mathcal{T}^*(t)$ be the time optimal trajectory for the Dub-L system. Further, suppose that the initial and the final conditions are such that the optimal control input for the laser is
\begin{align*}
    \omega^*(t) = 
    \begin{cases}
        \omega_M, \text{ if } t\in[0,\tau], \\
        0 \text{, otherwise,}
    \end{cases}
\end{align*}
for some time $\tau<t_f$. Then, for any $\epsilon>0$ such that $\tau+\epsilon\leq t_f$, there exists another optimal control input for the laser 
\begin{align*}
    \tilde{\omega}^*(t) = 
    \begin{cases}
        \omega_M, \text{ if } t\in[\epsilon,\tau+\epsilon], \\
        0 \text{, otherwise,}
    \end{cases}
\end{align*}
that ensures capture at the same time. Hence, we may have infinitely many candidates for the optimal trajectory.  In order to ascertain and characterize a unique class of trajectories between any two pair of initial and final configuration, we make the following assumption:

\begin{assumption}\label{assum:laser_on_once}
In a general setting, $\omega(0)=0$. A non-zero control can be applied to the laser at time $t_l\in[0,t_f]$. However, $\omega(t)\neq 0$, $\forall t\in [t_l,t_f] $.
\end{assumption}

Based on Assumption \ref{assum:laser_on_once}, we formulate this problem as a two stage optimal control problem with the following kinematic model:
\begin{align}\label{kinModSwitch}
    \dot{\textbf{x}} = 
    \begin{cases}
        f_1(\textbf{x},\textbf{u}), \text{ if } t<t_l,\\
        f_2(\textbf{x},\textbf{u}), \text{ otherwise},
    \end{cases}
\end{align}
where,
\begin{align*}
    &f_1(\textbf{x},\textbf{u}) = \begin{bmatrix} \cos(\theta) & \sin(\theta) & \frac{u}{\rho} & \frac{u}{\rho} \end{bmatrix}^\top,\\
    &f_2(\textbf{x},\textbf{u}) = \begin{bmatrix} \cos(\theta) & \sin(\theta) & \frac{u}{\rho} & \frac{u}{\rho}+\omega \end{bmatrix}^\top.
\end{align*}

We now formally describe the objective of this work.

\textbf{Problem Statement:} Given the kinematic equations in \eqref{kinModSwitch} and initial state vector $\textbf{x}_0$, determine the optimal control ${\bf u}^*\in U$, where $U=[-1,1]\times[-\omega_M,\omega_M]$, such that the terminal constraints in \eqref{eq:terminal_constraints} are satisfied in minimum time $t^*_f$. 

\section{Necessary Conditions}\label{sec:necessary}
In this section, we apply the Pontryagin maximum principle \cite{pontryagin1987mathematical} and characterize the optimal control for the Dub-L system. 

Let $\textbf{p}=\begin{bmatrix}p_x & p_y & p_{\theta} & p_{\psi}\end{bmatrix}^\top \in \mathbb{R}^4$ denote the costate of \textbf{x}. Then, given that the aim is to minimize the final time at which the target is captured,
we write the Hamiltonian $H$, as
\begin{align*}
H(\textbf{x},\textbf{u},\textbf{p},p_0) = \begin{cases} H_1, & \text{if } t\in [0,t_l), \\
H_2, & \text{otherwise,} \end{cases}
\end{align*}
where,
\begin{equation}
\begin{split}
    &H_1(\textbf{x},\textbf{u},\textbf{p},p_0) = p_0+p_x(t)\cos(\theta(t))+\\
    &p_y(t)\sin(\theta(t))+(p_{\theta}(t)+p_{\psi}(t))\frac{u(t)}{\rho},~\forall t\in [0,t_l),\\
    &H_2(\textbf{x},\textbf{u},\textbf{p},p_0) = p_0+p_x(t)\cos(\theta(t))+p_y(t)\sin(\theta(t))\\
    &+(p_{\theta}(t)+p_{\psi}(t))\frac{u(t)}{\rho} +p_{\psi}(t)\omega(t),~\forall t\in [t_l,t_f],
\end{split}
\end{equation}
and $p_0\geq 0$ is the abnormal multiplier. 
Although the costates may be discontinuous at time $t_l$, they must satisfy the adjoint equations in the time interval $[0,t_l)$ and $(t_l,t_f]$. The adjoint equations for $t\in[0,t_l)$ are
\begin{align*}
    &\dot{p}_x(t)=-\frac{\partial H_1}{\partial x}, \dot{p}_y(t)=-\frac{\partial H_1}{\partial y}, \dot{p}_{\theta}(t)=-\frac{\partial H_1}{\partial \theta}, \\
    &\dot{p}_{\psi}(t)=-\frac{\partial H_1}{\partial \psi}
\end{align*}
which yields that for $t\in[0,t_l)$
\begin{align*}
    &\dot{p}_x(t)=\dot{p}_y(t)=\dot{p}_{\psi}(t) = 0, \\
    &\dot{p}_{\theta}(t) = p_x(t)\sin(\theta)-p_y(t)\cos(\theta).
\end{align*}
Similarly, the adjoint equations for all $t\in(t_l,t_f]$ yields
\begin{align*}
    &\dot{p}_x(t)=\dot{p}_y(t)=\dot{p}_{\psi}(t) = 0, \\
    &\dot{p}_{\theta}(t) = p_x(t)\sin(\theta)-p_y(t)\cos(\theta).
\end{align*}
This implies that $p_x(t),p_y(t),$ and $p_{\psi}(t)$ are constant in the time interval $[0,t_l)$ and $(t_l,t_f]$. Let $t_l^-$ (resp. $t_l^+$) denote the time just before (resp. after) time $t_l$. Then, at time $t_l$, since there is no cost associated with switching from $f_1(\textbf{x},\textbf{u})$ to $f_2(\textbf{x},\textbf{u})$ and the state does not exhibit jumps at time instant $t_l$, we obtain \cite[Ch. 3]{bryson2018applied}
\begin{align}
    &p_x(t_l^-)=p_x(t_l^+), &p_y(t_l^-)=p_y(t_l^+), \nonumber \\ 
    &p_{\psi}(t_l^-)=p_{\psi}(t_l^+),
    &p_{\theta}(t_l^-)=p_{\theta} (t_l^+), \label{eq:p_tl} \\ &H_1(t_l^-)=H_2(t_l^+). \label{eq:H_tl}
\end{align}
Integrating $p_{x}(t),p_{y}(t),p_{\theta}(t),$ and $p_{\psi}(t)$ yields that $\forall t \in [0,t_f]$
\begin{subequations}
    \begin{equation}\label{eq:px_py_adjoint}
        p_x(t) = c_x,~~ p_y(t) = c_y,
    \end{equation}
    \begin{equation}\label{eq:p_psi_adjoint}
        p_{\psi}(t) = c_{\psi}
    \end{equation}
    \begin{equation}\label{eq:p_theta_adjoint}
        p_{\theta}(t)=c_xy(t)-c_yx(t)+c_0,
    \end{equation}
\end{subequations}
where $c_x,c_y,c_{\psi},$ and $c_0$ are some constants. From the Pontryagin maximum principle,
\begin{equation}\label{eq:H_zero}
  H(\textbf{x},\textbf{u},\textbf{p},p_0)\equiv 0, \forall t\in [0,t_f].
\end{equation}

Note that the terminal constraint in equation \eqref{eq:terminal_constraints_a} is an inequality constraint and hence the final position of the Dub-L system can lie inside the closed disk $\mathcal{C}$ defined as
\begin{align}
    \mathcal{C} := \{(x, y)\in\mathds{R}^2:~x^2+y^2-r^2 \leq 0\}.
\end{align}

Later in Theorems \ref{thm:on_C}, \ref{thm:CSC_on_C}, and \ref{thm:CC_on_C}, we will show that, except for in one trajectory, the final location will lie on $\partial C$, where $\partial C$ denotes the boundary of the set $\mathcal{C}$. Specifically, $\partial\mathcal{C}$ denotes the circle of radius $r$ centered at the origin. Therefore, we can now treat the constraint in equation $\eqref{eq:terminal_constraints_a}$ as an equality constraint and write the transversality conditions accordingly.
 
Let $\lambda_1\in \mathds{R}$ and $\lambda_2\in \mathds{R}$ be constant multipliers and let $g_1(\textbf{x})\coloneqq x^2(t_f)+y^2(t_f)-r^2$ and $g_2(\textbf{x}) \coloneqq \pi+\atantwo\left(y(t_f),x(t_f)\right) -\psi(t_f)$. Then the transversality conditions
\begin{align*}
    &p_x(t_f) = \lambda_1\frac{\partial g_1(\textbf{x})}{\partial x_f} + \lambda_2\frac{\partial g_2(\textbf{x})}{\partial x_f},\\
    &p_y(t_f) = \lambda_1\frac{\partial g_1(\textbf{x})}{\partial y_f} + \lambda_2\frac{\partial g_2(\textbf{x})}{\partial y_f},\\
    &p_{\theta}(t_f) = \lambda_1\frac{\partial g_1(\textbf{x})}{\partial \theta_f} + \lambda_2\frac{\partial g_2(\textbf{x})}{\partial \theta_f},\\
    &p_{\psi}(t_f) = \lambda_1\frac{\partial g_1(\textbf{x})}{\partial \psi_f} + \lambda_2\frac{\partial g_2(\textbf{x})}{\partial \psi_f},
\end{align*}
yields
\begin{subequations}\label{eq:transversality}
    \begin{equation}\label{eq:c_x_trans}
    p_x(t_f) = 2\lambda_1 x(t_f)-\lambda_2\frac{y(t_f)}{x^2(t_f)+y^2(t_f)},
    \end{equation}
    \begin{equation}\label{eq:c_y_trans}
        p_y(t_f) = 2\lambda_1 y(t_f) + \lambda_2\frac{x(t_f)}{x^2(t_f)+y^2(t_f)},
    \end{equation}
    \begin{equation}\label{eq:p_theta_trans}
        p_{\theta}(t_f) = 0,
    \end{equation}
    \begin{equation}\label{eq:p_psi_trans}
        p_{\psi}(t_f) = -\lambda_2.
    \end{equation}
\end{subequations}
From equation \eqref{eq:px_py_adjoint}, by substituting $p_x(t_f)=c_x$ and $p_y(t_f)=c_y$ in equation \eqref{eq:p_theta_adjoint} and using equations \eqref{eq:c_x_trans}, \eqref{eq:c_y_trans}, and \eqref{eq:p_theta_trans} yields
\begin{equation}\label{eq:lambda_2isc0}
    \lambda_2=c_0\implies c_{\psi}=-c_0.
\end{equation}
Note that since neither of the constraints defined in equation \eqref{eq:terminal_constraints} depends on the state $\theta$, equation \eqref{eq:p_theta_trans} holds even if the constraint specified in equation \eqref{eq:terminal_constraints_a} holds with inequality.



Further, from Pontryagin maximum principle, the optimal control $\textbf{u}^*$ is given as  $\argmin_{\textbf{u}\in U} H(\textbf{x},\textbf{u},\textbf{p},p_0)$ yielding
\begin{equation}\label{eq:u_opt_sing}
\begin{split}
    u^*(t) = \begin{cases}
-1,& p_{\theta}(t) + p_{\psi}(t) > 0\\
+1,& p_{\theta}(t) + p_{\psi}(t) < 0\\
\text{undetermined},& p_{\theta}(t) + p_{\psi}(t) = 0,
    \end{cases}
    \end{split}
\end{equation}
\begin{equation}\label{eq:w_opt_costates}
 \omega^*(t) = \begin{cases}
-\omega_M,& p_{\psi}(t) > 0\\
\omega_M,& p_{\psi}(t) < 0\\
\text{undetermined},& p_{\psi}(t)=0.
    \end{cases}
\end{equation}
The control $u(t)$ (resp. $\omega(t)$) when $p_{\theta}(t)+p_{\psi}(t)=0$ (resp. $p_{\psi}(t)=0$) is called \emph{singular control} and cannot be determined from $\argmin_{\textbf{u}\in U} H(\textbf{x},\textbf{u},\textbf{p},p_0)$.
In what follows, the locations at which the pose trajectory of the Dub-L system switches from one value of $u(t)$ to another value of $u(t)$ will be referred to as the \emph{switching points}.
We now determine the singular control $u^*(t)$ and $\omega^*(t)$. 

\begin{theorem}\label{thm:costates_0_switching}
    $p_{\theta}(t)+p_{\psi}(t)=0$ at the switching points as well as on the straight line segments of the pose trajectory of the Dub-L system.
\end{theorem}
\begin{proof}
    From equation \eqref{eq:p_psi_adjoint} and equation \eqref{eq:lambda_2isc0}, since $p_{\psi}(t)=c_{\psi}=-c_0$, $p_{\theta}(t)+p_{\psi}(t)= 0 \implies c_xy(t)-c_yx(t)=0$. If $c_x y(t)-c_y x(t)=0$ for any non-zero interval of time then, from elementary geometry, the trajectory is a straight line segment that lies on the line passing through the origin. Finally, as $u(t)$ changes sign at the switching point between two arcs and since $p_{\theta}(t)+p_{\psi}(t)$ is continuous, it follows that $p_{\theta}(t)+p_{\psi}(t)=0$ at the switching point between two arcs. 
\end{proof}
\medskip
From Theorem \ref{thm:costates_0_switching}, we obtain 
\begin{equation}\label{eq:u_opt_costates}
\begin{split}
    \argmin_{u\in[-1,1]} H_1 \Rightarrow u^* = 
    \begin{cases}
        -1, \;p_{\theta}(t)+p_{\psi}(t) > 0\\
        +1, \;p_{\theta}(t)+p_{\psi}(t) < 0\\
        ~~ 0, \;p_{\theta}(t)+p_{\psi}(t) = 0,
            \end{cases}
    \end{split}
\end{equation} 
From equation \eqref{eq:u_opt_costates}, it follows that the pose trajectory of the Dub-L system comprises arcs of radius $\rho$ (when $u(t)=+1$ or $u(t)=-1$) and straight line segments (when $u(t)=0$). As a circular arc can be a right turn or a left turn, in the sequel, a right turn circular arc will be denoted as $R$ and a left turn circular arc will be denoted as $L$. The following result follows immediately from Theorem \ref{thm:costates_0_switching}.
\medskip
\begin{corollary}\label{cor:switching}
    On any optimal pose trajectory, all switching points and all of the line segments are collinear with the target location.
\end{corollary}
\begin{proof}
    The result follows directly from Theorem \ref{thm:costates_0_switching}.
\end{proof}
\medskip
In what follows, since $p_{\psi}(t)=c_{\psi}$ is a constant, we will drop the dependence of time from $p_{\psi}(t)$. Further, we will use $\mathcal{D}_0$ to denote the line that passes through the switching points of the pose trajectory and the target. The next result characterizes that when $p_{\psi}=0$, the final location of the Dub-L system lies on $\partial \mathcal{C}$ and will be useful to establish the singular control of the laser.
\medskip
\begin{lemma}\label{lem:on_C}
    Let $\tau\in[t_0,t_f]$ denote the first time instant at which $x^2(\tau)+y^2(\tau)=r^2$ holds. Then, $t_f=\tau$ for an optimal trajectory for the Dub-L system if $p_{\psi}=0$. 
\end{lemma}
\begin{proof}
    From equation \eqref{eq:p_psi_adjoint} and equation \eqref{eq:p_psi_trans}, $p_{\psi}=0\implies \lambda_2=0$. This means that the optimal trajectory for the Dub-L system is identical to the optimal trajectory obtained without the constraint specified in equation \eqref{eq:terminal_constraints_b}. By simple geometrical arguments, it is trivial to verify that given the constraint defined in \eqref{eq:terminal_constraints_a}, the problem reduces to determining the optimal pose trajectory for the Dub-L system such that it ends at $\partial\mathcal{C}$ and the result follows.
\end{proof}
\medskip
\begin{figure}[t]
    \centering
    \includegraphics[scale=1.1]{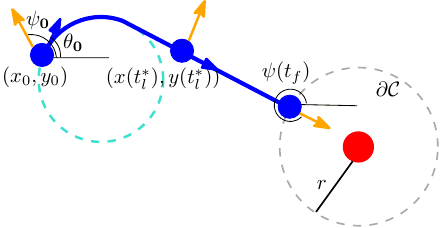}
    \caption{Illustration for proof of Theorem \ref{thm:singular_omega}. The red dot denotes the target and the gray dashed circle denotes $\partial\mathcal{C}$. $|\psi(t_l^*)-\psi(t_f)| > \pi$ and $\text{sgn}(\psi(t_l^*)-\psi(t_f))=-1$. Thus, $\omega^*=\omega_M$.}
    \label{fig:sing_omega}
\end{figure}

\begin{theorem}\label{thm:singular_omega}
    If $p_{\psi}=0$, then
    \begin{align*}
        \omega^* = 
        \begin{cases}
            \text{sgn}(\psi(t_l^*)-\psi(t_f))\omega_M, \text{ if } |\psi(t_l^*)-\psi(t_f)| \leq \pi\\
            -\text{sgn}(\psi(t_l^*)-\psi(t_f))\omega_M, \text{ otherwise.}
        \end{cases}
    \end{align*}
\end{theorem}
\smallskip
\begin{proof}
    From Lemma \ref{lem:on_C} and since $p_{\psi}=0\implies \lambda_2=0$, it follows that the laser has sufficient time to turn towards the final orientation while the vehicle moves to the final location. This concludes the proof.
\end{proof}

To summarize, in this section, we have expressed the necessary conditions arising from the Pontryagin maximum principle and have characterized the optimal control for the Dub-L system. These conditions will now be used to characterize the time optimal trajectory of the Dub-L system.



\section{Characterization of Optimal Trajectory}\label{sec:charac_shortest_path}

In this section, we characterize the minimum time trajectory for the Dub-L system. 
We start by establishing that the direction of the Dubins vehicle of the Dub-L system does not change at time $t=t_l$.
Note that if $t_l$ is such that the Dubins vehicle is at a switching point, then from equation \eqref{eq:u_opt_costates}, $u(t_l)=0$.
\medskip
\begin{lemma}\label{lem:concat}
    Suppose that the Dub-L system is not located at a switching point at time $t_l$. Then, $u(t_l^+)=u(t_l^-)$.
\end{lemma}
\begin{proof}
As $p_{\psi}$ is constant and, from equation \eqref{eq:p_tl}, $p_{\theta}(t_l^-)=p_{\theta}(t_l^+)$, it follows that $p_{\theta}(t_l^-)+p_{\psi}(t_l^-)=p_{\theta}(t_l^+)+p_{\psi}(t_l^+)$. The result then follows from equation \eqref{eq:u_opt_costates}.
\end{proof}
\medskip
Let, for some number $z\in\mathbb{R}$, $\text{sgn}(z)$ denote the sign function defined as
\begin{align*}
    \text{sgn}(z) = \begin{cases}
        -1, &\text{if } z<0,\\
        0, &\text{if } z=0,\\
        1, &\text{otherwise.}
    \end{cases}
\end{align*}
Then the next result establishes that, for any pose trajectory  of the Dub-L system that ends with an arc, the laser and the Dubins vehicle turn in the same direction in the final segment.
\medskip
\begin{figure}[t]
    \centering
    \includegraphics[scale=1.4]{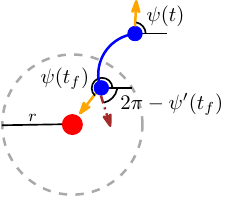}
    \caption{Illustration of trajectories $\mathcal{T}(t)$ and $\mathcal{T}'(t)$ for proof of Lemma \ref{lem:same_sign}. The red dot denotes the target and the gray dashed circle denotes $\partial \mathcal{C}$. The blue dot denotes the Dubins vehicle. Since the pose trajectory is same in both $\mathcal{T}(t)$ and $\mathcal{T}'(t)$, it is denoted by a single blue curve. The laser turns anti-clockwise (depicted in yellow color) in $\mathcal{T}(t)$ and clockwise (depicted in brown color) in $\mathcal{T}'(t)$. $\Delta\psi'>\Delta \psi$ holds.}
    \label{fig:same_sign}
\end{figure}
\begin{lemma}\label{lem:same_sign}
    Suppose that the optimal pose trajectory of the Dub-L system ends with a circular arc $C$. Then, $\text{sgn}(\omega^*(t))=\text{sgn}(u^*(t))$ in the final $C$ type segment.
\end{lemma}
\begin{proof}
    Let $\mathcal{T}(t):=\Gamma(t)|\mathcal{L}(t)$ denote a trajectory for the Dub-L system with the pose trajectory $\Gamma(t)$ such that it ends with a $C$ type segment and the laser's trajectory $\mathcal{L}(t)$ such that $\text{sgn}(\omega^*_{\mathcal{T}}(t))=-\text{sgn}(u^*(t))$ in the final $C$ segment of $\Gamma(t)$, where $\omega_{\mathcal{T}}(t)$ denotes the control of the laser in the final $C$ segment in trajectory $\mathcal{T}(t)$. 
    Now consider another trajectory $\mathcal{T}'(t):=\Gamma(t)|\mathcal{L}'(t)$ with the same pose trajectory $\Gamma(t)$ and the laser's trajectory $\mathcal{L}'(t)$ such that the laser turns clockwise in the final $C$ segment. Formally, $\mathcal{L}'(t)$ is such that $\text{sgn}(\omega^*_{\mathcal{T}'}(t))=\text{sgn}(u^*(t))$ in the final $C$ segment.
    Note that if $\Gamma(t)$ is just a $C$ segment, then $\text{sgn}(\omega^*_{\mathcal{T}'}(t))=\text{sgn}(u^*(t))$ in the entire trajectory $\mathcal{T}'(t)$. Further note that the constraint specified in equation \eqref{eq:terminal_constraints_b} may not hold for $\mathcal{T}'(t)$.
    Since the pose trajectory is same in both $\mathcal{T}(t)$ and $\mathcal{T}'(t)$, the time that the laser rotates in the final $C$ segment must be equal in both $\mathcal{T}(t)$ and $\mathcal{T}'(t)$. Mathematically,
    \begin{align*}
        \frac{\Delta\psi}{\omega_M-\tfrac{1}{\rho}} = \frac{\Delta\psi'}{\omega_M+\tfrac{1}{\rho}}\implies \Delta\psi'>\Delta \psi,
    \end{align*}
    where $\Delta\psi$ (resp. $\Delta\psi'$) denotes the angle that the laser rotates in the final $C$ segment of $\Gamma(t)$ (resp. $\Gamma'(t)$) and without loss of generality, we assumed that $\omega_M>\frac{1}{\rho}$. 
    Now, based on whether the final location of the dub-L system ends on $\partial\mathcal{C}$ or not in $\Gamma(t)$, there are two cases.

    \textbf{Case 1 ($x^2(t_f)+y^2(t_f)<r^2$):} The first case is that $(x(t_f),y(t_f))$ does not lie on $\partial\mathcal{C}$ (cf. Fig. \ref{fig:same_sign}). Since $\Delta\psi'>\Delta \psi$, it follows that by reducing the length of the final $C$ curve in the pose trajectory of $\mathcal{T}'(t)$ and, if required, having $\omega(t)=0$ for some non-zero interval of time, $\Delta\psi' = \Delta\psi$ can be achieved in $\mathcal{T}'(t)$. This implies that the time taken by the Dub-L system in $\mathcal{T}'(t)$ is less than that in $\Gamma(t)$ and thus, $\Gamma(t)$ cannot be optimal. 

    \textbf{Case 2 ($x^2(t_f)+y^2(t_f)=r^2$):} In this case, since the constraint specified in equation \eqref{eq:terminal_constraints_a} holds with equality, we can apply the transversality conditions specified in equation \eqref{eq:transversality}. In particular, without loss of generality, assume that at time $t_f$, $u^*(t_f)=-1$ and $\omega^*(t_f)=\omega_M$. Using equation \eqref{eq:p_theta_adjoint} and equation \eqref{eq:p_theta_trans}, it follows that $c_x y(t_f)-c_yx(t_f)=-c_0$. Since $u^*(t_f)=-1$, from equation \eqref{eq:u_opt_costates}, $c_x y(t_f)-c_yx(t_f) >0$ implying that $c_0<0$.
    This further implies, from equation \eqref{eq:w_opt_costates} and equation \eqref{eq:lambda_2isc0}, that $\omega^*(t_f)=-\omega_M$ which is a contradiction. Thus, $\text{sgn}(\omega^*(t_f))=\text{sgn}(u^*(t_f))$. Finally, since $c_{\psi}=-c_0$ is constant in $[0,t_f]$ and $u^*(t)$ does not change sign in the final $C$ segment, the result follows.
    
    Thus, we have shown that for any optimal trajectory with the pose trajectory that ends with a $C$ segment, $\text{sgn}(\omega^*(t))=\text{sgn}(u^*(t))$ holds. This concludes the proof.
\end{proof}
\medskip

In what follows, we will characterize the set of candidate trajectories for the Dub-L system. We will first characterize the abnormal solution and then characterize the set of trajectories for the Dub-L system when $t_l^*>0$ and when $t_l^*= 0$, separately. Finally, in the sequel, we will use $\phi\coloneqq \atantwo(c_y,c_x)$.

\subsection{Characterization of the abnormal solution ($p_0=0$)}

In this section, we characterize the time optimal trajectory when $p_0=0$. In particular, we will show that the time optimal pose trajectory consists of either only circular arcs or only straight line segments. Note that, under Assumption \ref{assum:laser_on_once}, the trajectory of the laser when $p_0=0$ is either clockwise or anti-clockwise.
\medskip
\begin{theorem}\label{thm:abnomral_S_bang}
    Let $p_0=0$.
    \begin{enumerate}
        \item If $\sqrt{c_x^2+c_y^2} + {c_{\psi}}\omega_M=0$, then $u(t)\equiv 0,~\forall t\in [t_0,t_f]$.
        \item Otherwise, the optimal pose trajectory consists only of circular arcs.
    \end{enumerate}
\end{theorem}
\begin{proof}
    We start with the proof of the first part. 
    If $\sqrt{c_x^2+c_y^2} +{c_{\psi}}\omega_M=0$ and $p_0=0$, then using equation \eqref{eq:H_zero}, we obtain that $\forall t\in[t_0,t_f]$
    \begin{multline*}
        \sqrt{c_x^2+c_y^2}(\cos(\theta(t)-\phi)-1)+(p_{\theta}+c_{\psi})\frac{u(t)}{\rho}=0.
    \end{multline*}
    From equation \eqref{eq:u_opt_costates}, the term $(p_{\theta}+c_{\psi})\frac{u(t)}{\rho}\leq 0$. Since both terms are non-positive, it follows that $\sqrt{c_x^2+c_y^2}(\cos(\theta-\phi)-1)=0$ and $(p_{\theta}+c_{\psi})\frac{u}{\rho}=0$ meaning that the path must be a straight line segment in $[t_0,t_f]$.

    For the second part, we proceed as follows. We will first show that $p_{\theta}+p_{\psi}$ solves the following differential equation:
    \begin{equation}\label{eq:differential}
        \dot{p}_{\theta}^2 + \left(-\frac{(p_{\theta}+c_{\psi})u}{\rho}-c_{\psi}\omega_M\right)^2 = c_x^2+c_y^2.
    \end{equation}    
    Then, we will establish that if $\sqrt{c_x^2+c_y^2} +{c_{\psi}}\omega_M\neq 0$, then the optimal pose trajectory consists of only circular arcs. 

    Recall that $\dot{p}_{\theta}(t) = c_x\sin(\theta)-c_y\cos(\theta)$. By dividing with $\sqrt{c_x^2+c_y^2}$ and substituting $\tfrac{c_x}{\sqrt{c_x^2+c_y^2}}$ (resp.  $\tfrac{c_y}{\sqrt{c_x^2+c_y^2}}$) as $\sin(\phi)$ (resp. $\cos(\phi)$), we obtain
    \begin{equation}\label{eq:pthetadot}
        \dot{p}_{\theta}^2 = (c_x^2+c_y^2)\sin^2(\theta-\phi)=(c_x^2+c_y^2)(1-\cos^2(\theta-\phi)).
    \end{equation}
    Using equation \eqref{eq:H_zero} yields $\sqrt{c_x^2+c_y^2}\cos(\theta-\phi)=-(p_{\theta}+c_{\psi})u/\rho-p_0-c_{\psi}\omega_M$. Substituting in equation \eqref{eq:pthetadot} and since $p_0=0$ yields equation \eqref{eq:differential}.
    
    Now, suppose that $u(t)\equiv 0$ over some interval $[\tau_1,\tau_2]$, $\tau_1\neq \tau_2$. This implies that $p_{\theta}(t)+c_{\psi}=0,~\forall t\in[\tau_1,\tau_2]$. If $p_0=0$, then equation \eqref{eq:differential} becomes
    \begin{align*}
        &\dot{p}_{\theta}^2 + (c_{\psi}\omega_M)^2 = c_x^2+c_y^2\\
        &\implies \dot{p}_{\theta}^2 =\left(\sqrt{c_x^2+c_y^2}+c_{\psi}\omega_M\right)\left(\sqrt{c_x^2+c_y^2}-c_{\psi}\omega_M\right).
    \end{align*}
    Since $\sqrt{c_x^2+c_y^2}+c_{\psi}\omega_M\neq 0$ and $c_{\psi}\omega_M\leq 0$, it follows that $\dot{p}_{\theta}\neq 0$ meaning that the singular control $u(t)\equiv 0$ over an interval of $[\tau_1,\tau_2]$ is not possible and the result follows. This concludes the proof. 
\end{proof}

Since we have now characterized trajectories when $p_0=0$, in what follows, we will consider that $p_0>0$ and normalize all costates by $p_0$.

Recall, from Section \ref{sec:introduction}, that when the orientation of the Dubins vehicle is not specified at the final location, then the problem of determining the time-optimal trajectory for the Dubins vehicle is known as the relaxed Dubins problem. In the next subsection, we will establish that determining the time-optimal trajectory for the Dub-L system is equivalent to a relaxed Dubins problem to a circle when $t_l^*>0$. 

\subsection{Characterization of Optimal Trajectory when $t_l^*>0$}\label{sec:charac_shortest_path_psi_0}
In this subsection we characterize the optimal trajectory for the Dub-L system when $t_l^*>0$.
We start with the following sufficient condition.
\medskip
\begin{lemma}\label{lem:t_l>0}
   If $t_l^*>0$, then $p_{\psi}=0$.
\end{lemma}
\begin{proof}
    Recall from equation \eqref{eq:H_tl}, that $H_1(t^-_l)=H_2(t^+_l)$. This yields
    \begin{align}\label{eq:relaxed_dubins}
   &1 + p_x \cos\theta(t^-_l) + p_y \sin\theta(t^-_l) + \left(p_\theta(t^-_l) + p_\psi\right) \frac{u(t^-_l)}{\rho} = \nonumber \\
   &1 + p_x \cos\theta(t^+_l) + p_y \sin\theta(t^+_l) + p_\theta(t^+_l) \frac{u(t^+_l)}{\rho} + \nonumber \\ 
   &p_\psi\left(\frac{u(t^+_l)}{\rho} + \omega(t^+_l)\right) \Rightarrow p_\psi \omega(t^+_l)=0,
\end{align}
where we used equation \eqref{eq:p_tl}. Given Assumption \ref{assum:laser_on_once}, since $\omega(t) \neq 0$ for $t\geq t_l$ , it follows that $p_{\psi}=0$. This concludes the proof. 
\end{proof}
\medskip
As a consequence of Lemma \ref{lem:t_l>0}, note that the laser's optimal control is characterized by Theorem \ref{thm:singular_omega} when $t_l^*>0$. 
\medskip

The following two lemmas will be instrumental in establishing the main result of this subsection. Recall, from Corollary \ref{cor:switching}, that the switching points and the straight line segments of the optimal pose trajectory are collinear and lie on the line $\mathcal{D}_0$. We start by establishing that the final location of the Dub-L system lies on $\mathcal{D}_0$ if $p_{\psi}=0$.
\medskip
\begin{lemma}\label{lem:final_loc_D0}
    Let $p_{\psi}=0$ and the pose trajectory for the Dub-L system be such that it consists of at least one switching point or a straight line segment. Then $(x(t_f),y(t_f))$ lies on the line $\mathcal{D}_0$.
\end{lemma}
\begin{proof} 
    Recall from equation \eqref{eq:p_theta_adjoint} that, for all $t\in [0,t_f]$, $p_{\theta}(t)=c_xy(t)-c_yx(t)+c_0$. Using equation \eqref{eq:p_theta_trans}, we obtain $c_xy(t_f)-c_yx(t_f)+c_0=0$. Now, from equation \eqref{eq:lambda_2isc0}, since $c_{\psi}=-c_0$ and using $c_{\psi}=0$ yields $c_xy(t_f)-c_yx(t_f)=0$ and the result follows.
\end{proof}
\medskip
\begin{lemma}\label{lem:no_CC}
    If $p_{\psi}=0$, then a pose trajectory of type $CC$ is not optimal.
\end{lemma}
\begin{proof}
    We defer the proof to the Appendix.
\end{proof}
\medskip
We now present the main result of this subsection.
\medskip
\begin{theorem}\label{thm:no_CC}
    The optimal pose trajectory for the Dub-L system is of type $CS$ or a subsegment of it if $p_{\psi}=0$.
\end{theorem}
\begin{proof}
    Lemma \ref{lem:no_CC} establishes that a pose trajectory of type $CC$ is not optimal. Thus, in this proof, we show that an optimal pose trajectory cannot be of type $CSC$ if $p_{\psi}=0$.
    
    Consider a pose trajectory of type $CSC$. From Lemma \ref{lem:final_loc_D0}, it follows that the final $C$ segment must be either of length $0$ or $2\pi\rho$. From Lemma \ref{lem:on_C}, a $C$ segment of length $2\pi\rho$ is not optimal. Thus, any pose trajectory of type $CSC$ is not optimal and the result is established.
\end{proof}


\medskip
We now conclude this section with the following remarks.
\medskip
\begin{remark}\label{rem:relaxed}
    Since $p_{\psi}=0 \implies\lambda_2=0$, the problem is equivalent to finding the time optimal trajectory of the relaxed Dubins problem to a circle\footnote{To the best of our knowledge, the relaxed Dubins problem to a circle has not been considered before.}. Thus, Theorem \ref{thm:no_CC} characterizes the set of optimal trajectories for relaxed Dubins problem to a circle as well. 
\end{remark}

\begin{remark}\label{rem:psi0}
    Lemma \ref{lem:t_l>0} characterizes a sufficient condition that the costate $p_{\psi}=0$ when $t_l^*>0$. However, it is possible that $p_{\psi}=0$ even when $t_l^*=0$.
    Thus, the results in Lemma \ref{lem:final_loc_D0}, Lemma \ref{lem:no_CC}, and Theorem \ref{thm:no_CC} will hold even when $t_l^*=0$ ( as long as $p_{\psi}=0$).
\end{remark}

To sum up, we characterized the time-optimal trajectory for the Dub-L system when the speed of the laser is such that it has sufficient time to turn towards the final orientation. In the next subsection, we consider the case when $t_l^*= 0$. Since, in this section, we already characterized trajectories with $p_{\psi}=0$ (cf. Remark \ref{rem:psi0}), we will consider that $p_{\psi}\neq 0$. 

\subsection{Characterization of Optimal Trajectory when $t_l^*= 0$}\label{sec:charac_shortest_path_psi_not0} 
In this section we will establish that the optimal pose trajectory for the Dub-L system is either of type $CSC$ or of type $CC$ or a subsegment of these two.
We start by establishing that a pose trajectory of type $CCC$ is not optimal.
\medskip
\begin{theorem}\label{thm:no_CCC}
    A pose trajectory of type $CCC$ for the Dub-L system is not optimal.
\end{theorem}
\begin{proof}
    We defer the proof to the Appendix.
\end{proof}
\medskip
\begin{remark}\label{rem:no_CCC_abnormal}
    Theorem \ref{thm:no_CCC} holds even when $p_0=0$. Thus, from Theorem \ref{thm:abnomral_S_bang}, the time-optimal pose trajectory for the Dub-L system is of type $S$, $C$, or $CC$ when $p_0=0$.
\end{remark}
\medskip
We highlight that the abnormal solution of the Dubins vehicle can be either a $C$ or a $CC$ type segment \cite{kaya2017markov}. For the Dub-L system, the abnormal solution of the pose trajectory is of type $S$, $C$, or $CC$. We now establish the main result of this work, i.e., the set of candidate pose trajectories for the Dub-L system.
\medskip
\begin{theorem}\label{thm:opt}
An optimal pose trajectory for the Dub-L system is either of type $CSC$ or $CC$ or a subsegment of these two.
\end{theorem}
\begin{proof}
    If an optimal pose trajectory of the Dub-L system contains a straight line segment, it follows from Corollary \ref{cor:switching} that the trajectory must be of type $CSC$ or a subsegment of it. Next, from Theorem \ref{thm:no_CCC}, it follows that the optimal path of the Dub-L system is either of type $CSC$ or $CC$ or a subsegment of it. 
    Finally, from Theorem \ref{thm:abnomral_S_bang}, the abnormal solutions are a subsegment of either a $CSC$ type trajectory or a $CC$ type trajectory. This concludes the proof.
\end{proof}
\medskip
Theorem \ref{thm:opt} characterizes the set of candidates of optimal pose trajectories for the Dub-L system. Combining these with the trajectory of the laser, i.e., clockwise or anti-clockwise, yields the set of optimal trajectories for the Dub-L system. Note that by considering all possible combinations of the laser trajectories with the set of possible pose trajectories, there are a total of $26$ candidate trajectories for the Dub-L system. However, from Lemma \ref{lem:same_sign}, the total number of candidate trajectories reduces to a total of $16$.
Table \ref{tab:traj} summarizes the set of candidate trajectories for the Dub-L system depending on the various conditions characterized until now. 

So far, we have established the set of candidate trajectories for the Dub-L system and characterized properties of the shortest trajectory. Next, we will establish that all of the pose trajectories for the Dub-L system that consists of either at least one switching point or a straight line segment ends on $\partial \mathcal{C}$.

\subsection{Conversion of inequality constraint to equality constraint}\label{sec:inequaltoequal}
In this section, we establish that all of the possible optimal pose trajectories for the Dub-L system, excluding a $C$ type trajectory, will end at a distance $r$ from the target. We will first establish that all pose trajectories for the Dub-L system end on $\partial \mathcal{C}$ if $t_l^*>0$. Then, we will establish that all pose trajectories (excluding a $C$ type) end on $\partial \mathcal{C}$ when $t_l^*=0$.

\begin{theorem}\label{thm:on_C}
    If $t_l^*>0$, then $x^2(t_f)+y^2(t_f)=r^2$ holds.
\end{theorem}
\begin{proof}
    The proof is straightforward. Suppose that $x^2(t_f)+y^2(t_f)<r^2$ holds. Since $t_l^*>0$, from Assumption \ref{assum:laser_on_once}, it follows that $\omega(t)=0$, $\forall t\in [0,t_l)$. Since $t_l^*>0$ then, by applying the laser's control $\omega(t)=\omega_M$, $\forall t\in [t_l^*-\epsilon,t_f]$, the length of the trajectory can be reduced while ensuring that the constraint in \eqref{eq:terminal_constraints_b} remain satisfied. This implies that the trajectory in consideration is not optimal and the result is established.
\end{proof}


We now establish that all of the pose trajectories, excluding a $C$ type, ends on $\partial \mathcal{C}$ if $t_l^*=0$. We will first establish the result for all pose trajectories that consists of a straight line segment followed by the pose trajectories that consists of only circular arcs.
\medskip
\begin{theorem}\label{thm:CSC_on_C}
    For any optimal pose trajectory for the Dub-L system that consists of a straight line, $x^2(t_f)+y^2(t_f)=r^2$ holds at time $t_f$.
\end{theorem}
\begin{proof}
    We defer the proof to the Appendix.
\end{proof}
\medskip
\begin{theorem}\label{thm:CC_on_C}
    For any pose trajectory of the Dub-L system that consists of only circular arcs and at least one switching point, $x^2(t_f)+y^2(t_f)=r^2$ holds at time $t_f$.
\end{theorem}
\begin{proof}
    The proof is omitted for brevity as it is analogous to that of Theorem \ref{thm:CSC_on_C}.
\end{proof}
\medskip

\begin{table}[t]
\caption{Trajectories for the Dub-L system. "$+$" denotes clockwise and "$-$" denotes anti-clockwise direction. "$|$" separates the pose trajectory and the laser trajectory of the Dub-L system.}
\begin{center}
\begin{tabular}{|p{75pt}|p{150pt}|p{115pt}|}
\hline
Condition on co-states& 
Possible candidates\\[0.5ex]
\hline
$p_{\psi}=0$ and $p_0\neq 0$& 
$RS|+$, $RS|-$, $LS|+$, $LS|-$, $R|+$, $L|-$, $S|+$, $S|-$\\
$p_{\psi}\neq 0$ and $p_0\neq 0$& 
$RSR|+$, $RSL|-$, $LSL|-$ $LSR|+$, $RL|-$, $LR|+$, $RS|+$, $RS|-$, $LS|+$, $LS|-$, $SR|+$, $SL|-$, $R|+$, $L|-$, $S|+$, $S|-$\\
$p_0=0$& 
$RL|-$, $LR|+$, $R|+$, $L|-$, $S|+$, $S|-$\\
\hline
\end{tabular}
\end{center}
\label{tab:traj}
\end{table}

Next, we provide a solution for the optimal trajectory. In particular, we will describe the procedure to obtain the complete solution for the optimal trajectory when $p_{\psi}=0$. When $p_{\psi}\neq 0$, we will establish that the solution can be obtained by solving a set of nonlinear equations.

\section{Solution for Optimal Trajectory}\label{sec:sol}
We now provide a complete parameterization of the trajectories of the Dub-L system characterized in the previous section. We will first describe the process to obtain the solution of the trajectories when $p_{\psi}=0$. Then, we will provide a semi-analytical solution of the trajectories when $p_{\psi}\neq 0$. Note that we will denote $t_1$ and $t_2$ as the time when the Dub-L system is at the first and the second switching point, respectively.

\subsection{Solution for trajectories when $t_l^*>0$}\label{subsec:sol_relaxed}
Recall, from Lemma \ref{lem:on_C}, that the final location of the Dub-L system lies on $\partial\mathcal{C}$ and the pose trajectory is of type $CS$ or a subsegment of it. Here, we only describe the solution procedure of $CS$ type pose trajectory since the solution procedure for its subsegments is analogous.

\begin{figure}[t]
    \centering
    \includegraphics[scale=1]{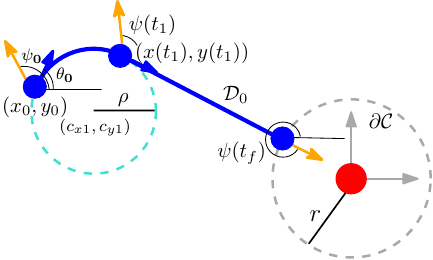}
    \caption{An illustration of $RS$ type trajectory when $t_l^*>0$ with the laser turning clockwise.}
    \label{fig:Traj_sol_CS}
\end{figure}

Given the initial state of the Dub-L system (cf. Fig. \ref{fig:Traj_sol_CS}), the center $(c_{x1},c_{y1})$ of the circle formed by the $C$ segment of the $CS$ pose trajectory is $(x_0+\rho\sin(\theta_0), y_0-\rho\cos(\theta_0))$. Note that Figure \ref{fig:Traj_sol_CS} is only an illustration and the description for an $LS$ trajectory is analogous. First, given the constraint in equation \eqref{eq:terminal_constraints_b} and since the pose trajectory is of type $CS$, it follows that $\psi(t_f)=\theta(t_f)$.
Then, from Corollary \ref{cor:switching}, as the $S$ segment is collinear with the target location and since the orientation of the vehicle does not change in the $S$ segment, $\theta(t_f)=\theta(t_1)$. Thus, $\theta(t_f)$ is determined by finding the distance from the center $(c_{x1},c_{y1})$ to the $S$ segment at the switching point. Mathematically,
\begin{align*}
    &\frac{|c_{y1}-\tan(\theta(t_f))c_{x1}|}{\sqrt{\tan^2(\theta(t_f))+1}}=\rho,\\
    &\implies \tan(\theta(t_f)) = \frac{c_{y1}c_{x1}\pm \rho\sqrt{c_{y1}^2+c_{x1}^2-1}}{c_{x1}^2-\rho^2}.
\end{align*}
Note that, for a specified $\theta_0$, one out of the two values of $\theta(t_f)$ can be eliminated.
Finally, by determining the intersection between the $S$ segment and $\partial\mathcal{C}$, we obtain $(x(t_f),y(t_f))$ as $\left(\frac{\pm r}{\sqrt{1+\tan^2(\theta(t_f))}}, \frac{\pm \tan(\theta(t_f))r}{\sqrt{1+\tan^2(\theta(t_f))}}\right)$. We now determine $t_l^*$.

Let $T_D$ denote the time taken by the Dub-L system to move from $(x_0,y_0)$ with orientation $\theta_0$ to the location $(x(t_f),y(t_f))$ with orientation $\theta(t_f)$. As location $(x(t_1), y(t_1))$ can be determined from elementary geometry and once $x(t_f), y(t_f),$ and $\theta(t_f)$ are determined, $T_D$ can be easily computed for a given pose trajectory.
Suppose that the laser turns clockwise as illustrated in Figure \ref{fig:Traj_sol_CS}. 

For a $CS$ type pose trajectory, since the angular speed of the laser changes depending on whether $t_l^*\geq t_1$ or $t_l^*< t_1$, $t_l^*$ is determined as follows.


Let $\alpha$ denote the angle of the $C$ segment and let $T_D^S\coloneqq \sqrt{(x(t_f)-x(t_1))^2+(y(t_f)-y(t_1))^2}$. Then, we first check if $t_l^*\geq t_1$. This is achieved in two steps. First, suppose that $t_l^*\geq t_1$ holds. Then, by equating the time taken by the laser to turn in the $C$ segment to the time taken by the Dubins vehicle yields
\begin{align*}
    \frac{\psi(t_1)-\psi_0}{u^*/\rho}=\rho\alpha\implies \psi(t_1) = \psi_0+u^*\alpha,
\end{align*}
where we use the fact that the laser is turning clockwise.
Once $\psi(t_1)$ is determined, the second step is to check if $T_D^S\geq \frac{\psi(t_f)-\psi(t_1)}{\omega^*}$ holds. Note that $u^*$ and $\omega^*$ are known for a given trajectory.
If $T_D^S\geq \frac{\psi(t_f)-\psi(t_1)}{\omega^*}$ holds, then it follows that $t_l^*\geq t_1$. Thus, in this case, $t_l^* = T_D^S-\frac{\psi(t_f)-\psi_0-u^*\alpha}{\omega^*}$.

Otherwise, i.e., if $T_D^S< \frac{\psi(t_f)-\psi(t_1)}{\omega^*}$ holds then, it follows that $t_l^*<t_1$. Analogous to the previous case, we first determine $\psi(t_l^*)$ from the following equation
\begin{align*}
    \frac{\psi(t_l^*)-\psi_0}{\tfrac{u^*}{\rho}} + \frac{\psi(t_f)-\omega^*T_D^S-\psi(t_l^*)}{\omega^*+\tfrac{u^*}{\rho}} = \rho\alpha,
\end{align*}
where we used the fact that $T_D^S = \frac{\psi(t_f)-\psi(t_1)}{\omega^*}$. Then, $t_l^* = \frac{\psi(t_l^*)-\psi_0}{u^*/\rho}$.

Note that, if the pose trajectory of the Dub-L system is of type $C$, then $t_l^*$ can be determined from the following equation
\begin{align*}
    \frac{\psi(t_l^*)-\psi_0}{\tfrac{u^*}{\rho}} + \frac{\psi(t_f)-\psi(t_l^*)}{\omega^*+\tfrac{u^*}{\rho}} = T_D.
\end{align*}
Similarly, for an $S$ type pose trajectory, $t_l^*$ can be computed from $\frac{\psi(t_f)-\psi(t_l^*)}{\omega^*} = T_D$.

\subsection{Solution for trajectories when $t_l^*= 0$}\label{subsec:t_l_0}
From Theorems \ref{thm:CSC_on_C} and \ref{thm:CC_on_C}, as the condition $x^2(t_f)+y^2(t_f)=r^2$ holds for both $CSC$ and $CC$ trajectory, we have that $x(t_f)=r\cos(\eta)$ and $y(t_f)=r\sin(\eta)$, where $\eta$ denotes the angle of the final location of the Dub-L system measured from the positive $x$-axis. Further, from equation \eqref{eq:c_x_trans} and equation \eqref{eq:c_y_trans},
\begin{equation}\label{eq:trans_rewrite}
    c_x=\lambda_1 r\cos(\eta)-\frac{c_0\sin(\eta)}{r}, c_y=\lambda_1 r\sin(\eta)+\frac{c_0\cos(\eta)}{r}.
\end{equation}
We now establish the solution for the $CC$ type trajectory followed by the $CSC$ type trajectory.

\begin{figure}[t]
    \centering
    \includegraphics[scale=1]{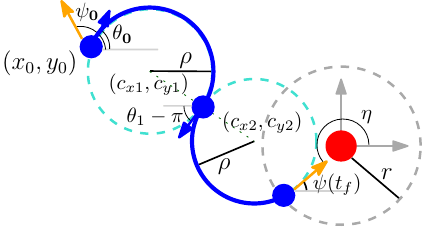}
    \caption{An illustration of $RL$ type trajectory when $t_l^*=0$.}
    \label{fig:traj_CC}
\end{figure}
\subsubsection{$CC$ type trajectory}
Let $(c_{x1},c_{y1})$ and $(c_{x2},c_{y2})$ denote the centers of the two circles of the $CC$ trajectory (cf. Fig. \ref{fig:traj_CC}). 
Then, as $\theta(t_1)$ is perpendicular to the line joining the two centers $(c_{x1},c_{y1})$ and $(c_{x2},c_{y2})$, we obtain
\begin{align*}
    \tan(\theta(t_1)) = -\frac{c_{x2}-c_{x1}}{c_{y2}-c_{y1}},
\end{align*}
where $(c_{x1},c_{y1})$ is $(x_0+\rho\sin(\theta_0), y_0-\rho\cos(\theta_0))$ and $(c_{x2},c_{y2})$ is $(x(t_f)-\rho\sin(\theta(t_f)), y(t_f)+\rho\cos(\theta(t_f)))$. Further,
$(x(t_1),y(t_1))$ can be written as $(c_{x2}+\rho\sin(\theta(t_1)),c_{y2}-\rho\cos(\theta(t_1)))$. Since $x(t_f) = r\cos(\eta)$ and $y(t_f)=r\sin(\eta)$, it follows that $x(t_1)$, $y(t_1)$, and $\theta(t_1)$ can be determined from $\eta$ and $\theta(t_f)$.
Note that, given a trajectory such as in Figure \ref{fig:traj_CC}, $u^*$ and $\omega^*$ are known.
Next, using equation \eqref{eq:H_zero} at $t=0$, $t=t_1$, and $t=t_f$ yields

\begin{multline}\label{eq:Hi_CC}
    H(0)=1+c_x\left(\cos(\theta_0)+\frac{y_0u_0^*}{\rho}\right)+c_y\left(\sin(\theta_0)-\frac{x_0u_0^*}{\rho}\right)\\
    +c_{\psi}\omega^*=0,
\end{multline}
\begin{multline}\label{eq:H1_CC}
H(t_1)=1+c_x\left(\cos(\theta(t_1))+\frac{y(t_1)\sin(\theta(t_1))}{x(t_1)}\right)\\
+c_{\psi}\omega^*=0,
\end{multline}
\begin{equation}\label{eq:H_f_CC}
    H(t_f)=1+c_x\cos(\theta(t_f))+c_y\sin(\theta(t_f))+c_{\psi}\left(\frac{u_f^*}{\rho}+\omega^*\right)=0,
\end{equation}
where we have used equation \eqref{eq:p_theta_trans} and Theorem \ref{thm:costates_0_switching}. Next, from Figure \ref{fig:traj_CC} and Lemma \ref{lem:same_sign}, as the vehicle turns anti-clockwise in the final $C$ segment, the laser must turn anti-clockwise during the entire trajectory. Finally, let $T_D=\rho(\alpha+\beta)$ be the time taken by the Dub-L system to move from the initial location and orientation to the final location and orientation, where $\alpha$ and $\beta$ denotes the angle subtended by the first and the second $C$ segment, respectively. Let $d_1$ denote the distance between $(x(t_1),y(t_1))$ and $(x(0),y(0))$. Then, from law of cosines, we obtain
\begin{align*}
    \cos(2\pi-\alpha) = \frac{2\rho^2-d_1^2}{2\rho}.
\end{align*}
Since $x(t_f) = r\cos(\eta)$ and $y(t_f)=r\sin(\eta)$, angle $\alpha$ (and analogously $\beta$), can be expressed as a function of $\eta$ and $\theta(t_f)$. This implies that $T_D$ can determined if $\eta$ and $\theta(t_f)$ can be determined.

Finally, let $T_L = \frac{\psi_0-\psi(t_1)}{\omega^*-\tfrac{1}{\rho}} + \frac{\psi(t_1)-\psi(t_f)}{\omega^*+\tfrac{1}{\rho}}$ be the time taken by the laser to orient to the final orientation $\psi(t_f)=\eta+\pi$. Using that $\rho\alpha = \frac{\psi_0-\psi(t_1)}{\omega^*-\tfrac{1}{\rho}}$, the orientation of the laser $\psi(t_1)$ can be expressed as a function of $\theta(t_f)$ and $\eta(t_f)$ implying that $T_L$ can be determined from $\eta$ and $\theta(t_f)$.
Then, using the fact that the time taken by the laser and the Dubins vehicle is equal yields
\begin{equation}\label{eq:time_equal}
    T_D(\eta,\theta(t_f))-T_L(\eta,\theta(t_f))=0.
\end{equation}
Thus, substituting $c_x$ and $c_y$ from equation \eqref{eq:trans_rewrite} in  equations \eqref{eq:Hi_CC}-\eqref{eq:time_equal} yields a set of four equations with four unknowns $\eta,\lambda_1,\theta(t_f)$, and $c_0$. Solving these equations yield $\theta(t_f),\psi(t_f),x(t_f),y(t_f)$.

\begin{figure}[t]
    \centering
    \includegraphics[scale=0.9]{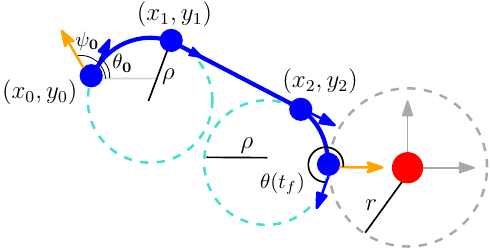}
    \caption{An illustration of $RSR$ type trajectory when $t_l^*=0$.}
    \label{fig:traj_CSC}
\end{figure}

\subsubsection{$CSC$ type trajectory}
We now characterize the solution of the $CSC$ trajectory of the Dub-L system. 

Analogous to the $CS$ trajectory in subsection \ref{subsec:sol_relaxed}, the first switching point and the orientation of the Dub-L system at the first switching point can be determined through elementary geometry. Then, using equation \eqref{eq:H_zero} at $t=0$, $t=t_1$, and $t=t_f$ yields

\begin{multline}\label{eq:H_i}
    H(0)=1+c_x\left(\cos(\theta_0)+\frac{y_0u_0^*}{\rho}\right)+c_y\left(\sin(\theta_0)-\frac{x_0u_0^*}{\rho}\right)+\\
    c_{\psi}\omega^*=0
\end{multline}
\begin{multline}
H(t_1)=1+c_x\left(\cos(\theta(t_1))+\frac{y(t_1)\sin(\theta(t_1))}{x(t_1)}\right)\\
+c_{\psi}\omega^*=0.
\end{multline}
\begin{multline}
        H(t_2)=1+c_x\left(\cos(\theta(t_1))+\frac{y(t_2)\sin(\theta(t_1))}{x(t_2)}\right)\\
        +c_{\psi}\omega^*=0.
\end{multline}
\begin{equation}\label{eq:H_f}
    H_f=1+c_x\cos(\theta(t_f))+c_y\sin(\theta(t_f))+c_{\psi}\left(\frac{u_f^*}{\rho}+\omega^*\right)=0
\end{equation}
where we have used equation \eqref{eq:p_theta_trans}, Theorem \ref{thm:costates_0_switching} and the fact that $\theta(t_1)=\theta(t_2)$ and $u^*$ and $\omega^*$ are known for a given trajectory (Fig. \ref{fig:traj_CSC}). Further, location $(x(t_2),y(t_2))$ can be expressed as $(r\cos(\eta)-\rho\sin(\theta(t_f))+\rho\sin(\theta(t_1)),r\sin(\eta)+\rho\cos(\theta(t_f))-\rho\cos(\theta(t_1)))$. Finally, substituting $c_x$ and $c_y$ from equation \eqref{eq:trans_rewrite} in equations \eqref{eq:H_i}-\eqref{eq:H_f} yields a system of four equations with four unknowns ($\eta$, $\lambda_1$, $\theta(t_f)$, and $c_0$), solving which yields $\theta(t_f),\psi(t_f),x(t_f),y(t_f)$.

It must be highlighted that if multiple solutions of $x(t_f),y(t_f)$, $\theta(t_f)$, and $\psi(t_f)$  are obtained by solving equations \eqref{eq:H_i}-\eqref{eq:H_f} then the solution that yields the trajectory with minimum time must be used.

\section{Numerical Simulations}\label{sec:numerics}
In this section, we present some numerical simulations to illustrate the properties characterized in this work. For all of our simulations, $r=\rho=1$.

\begin{figure}[t]
    \centering
    \includegraphics[scale=0.39]{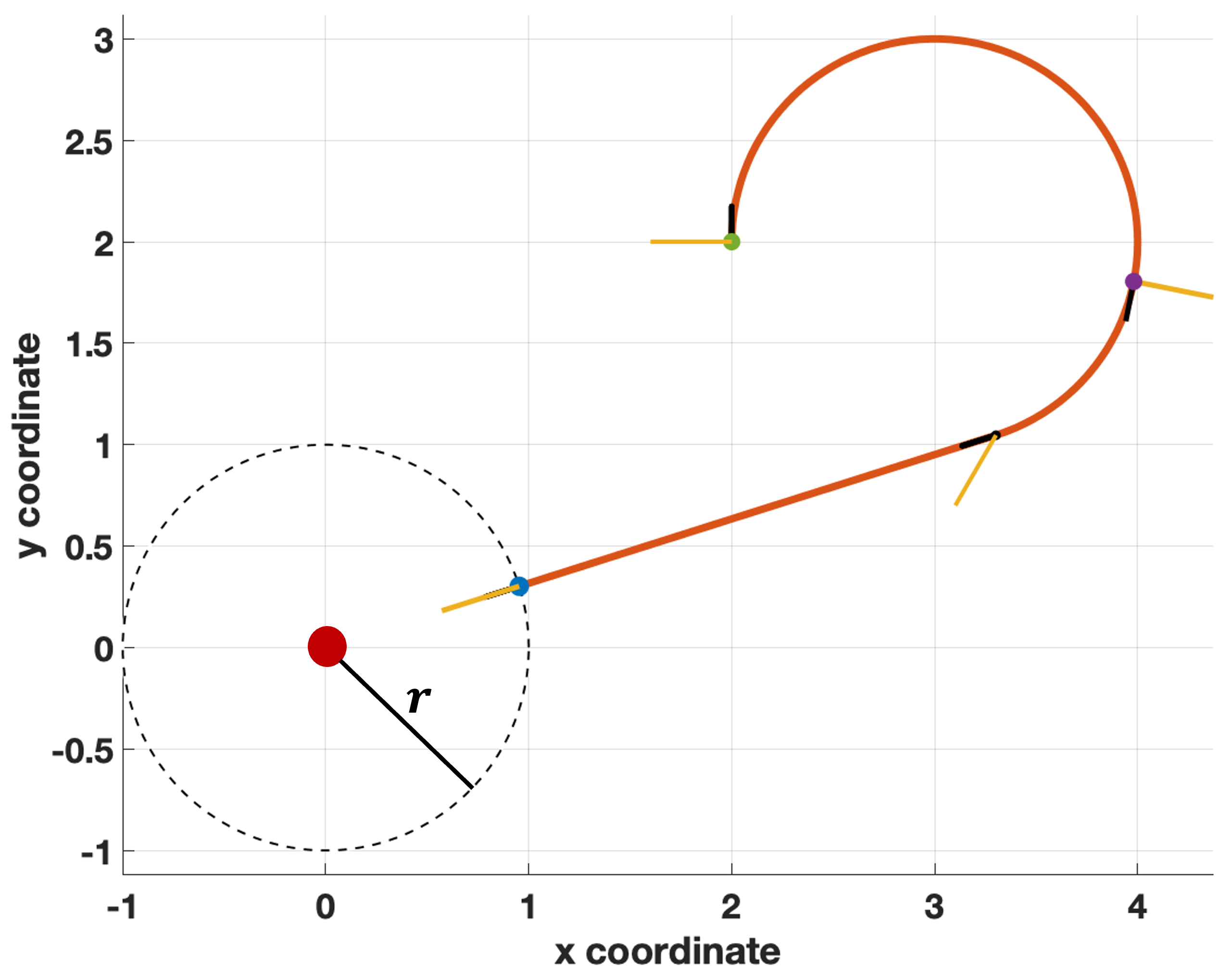}
    \caption{A $CS$ type trajectory when $t_l^*>0$ with $x_0,y_0,\theta_0,\psi_0$, and $\omega_M$ as $2,2,\frac{\pi}{2},\pi$, and $0.3$ respectively. The black dashed circle denotes $\partial\mathcal{C}$. The initial location is denoted as a green dot and the final location is denoted as the blue dot. The yellow line denotes the laser. The location $\left(x(t_l^*),y(t_l^*)\right)$ is denoted as purple dot.}
    \label{fig:CS_t_no0}
\end{figure}

Figure \ref{fig:CS_t_no0} depicts the shortest trajectory when $t_l^*>0$, where the final location and the orientation of the Dub-L system is determined according to Subsection~\ref{subsec:sol_relaxed}. The initial state, i.e., $x_0,y_0,\theta_0,\psi_0$, and $\omega_M$ are set to $2,2,\frac{\pi}{2},\pi$, and $0.3$, respectively. It is evident that the condition $x^2(t_f)+y^2(t_f)=r^2$ holds and the switching point and the straight line segment are collinear with the target location. Further, the values of $T_D$ and $t_l^*$ are determined to be $6.86$ and $2.948$, respectivley.

\begin{figure}[t]
    \centering
    \includegraphics[scale=0.4]{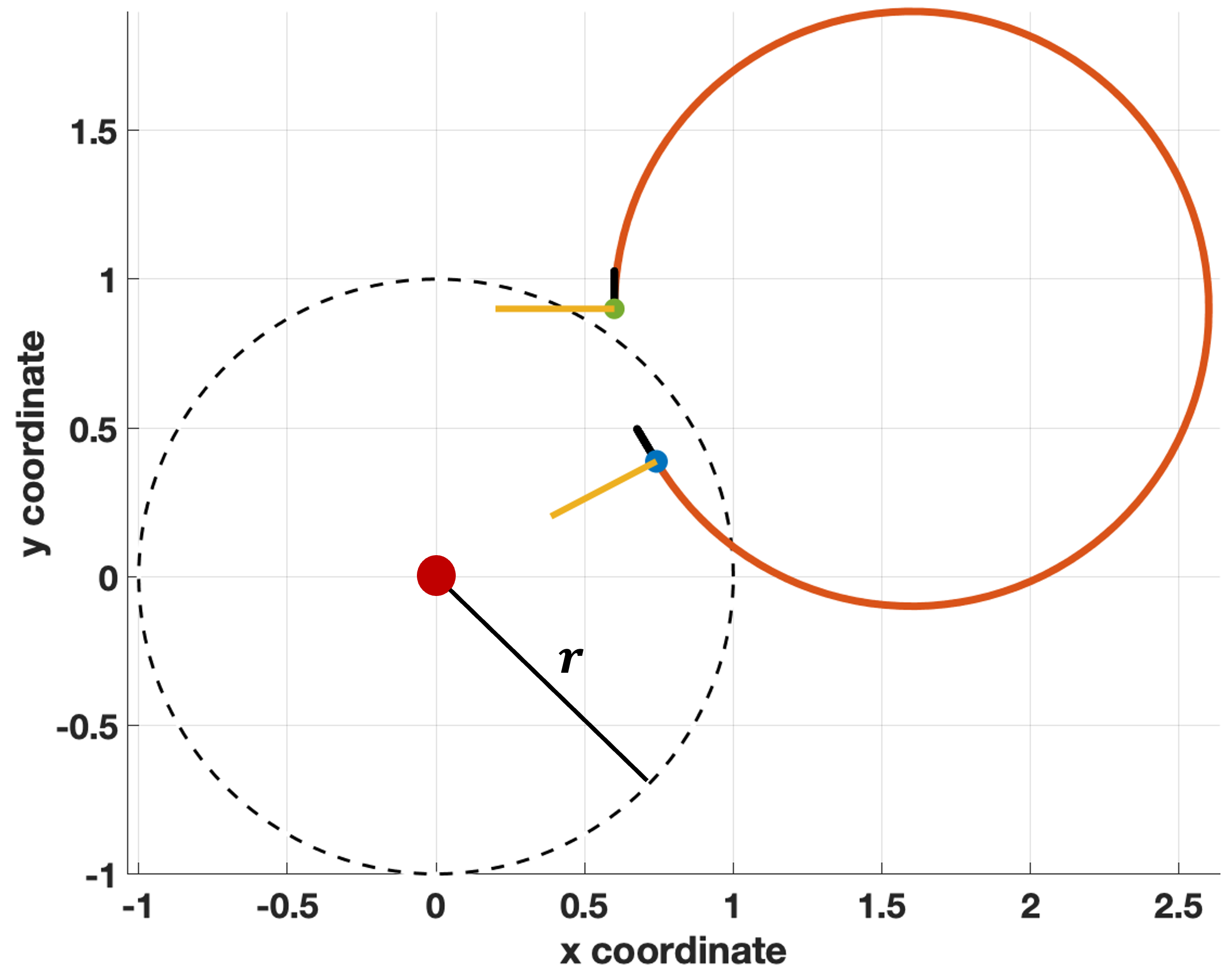}
    \caption{A $C$ type trajectory when $t_l^*=0$ with $x_0,y_0,\theta_0,\psi_0$, and $\omega_M$ as $0.6,0.9,\frac{\pi}{2},\pi$, and $0.01$ respectively. The black dashed circle denotes $\partial\mathcal{C}$. The initial location is denoted as a green dot and the final location is denoted as the blue dot. The yellow line denotes the laser and the black line denotes the orientatio of the Dubins vehicle.}
    \label{fig:C_num}
\end{figure}

Figure \ref{fig:C_num} shows a $C$ type trajectory with $t_l^*=0$ with $x_0,y_0,\theta_0,\psi_0$, and $\omega_M$ as $0.5,0.5,\frac{\pi}{2},\pi$, and $0.01$, respectively. Recall that a $C$ type trajectory is the only trajectory which does not end at $\partial\mathcal{C}$. To determine the final location and the orientation of the Dub-L system, we use \emph{fsolve} function in MATLAB to obtain the solution of equations \eqref{eq:Hi_CC}-\eqref{eq:time_equal} characterized in Subsection \ref{subsec:t_l_0}. 

\begin{figure}[t]
    \centering
    \includegraphics[scale=0.39]{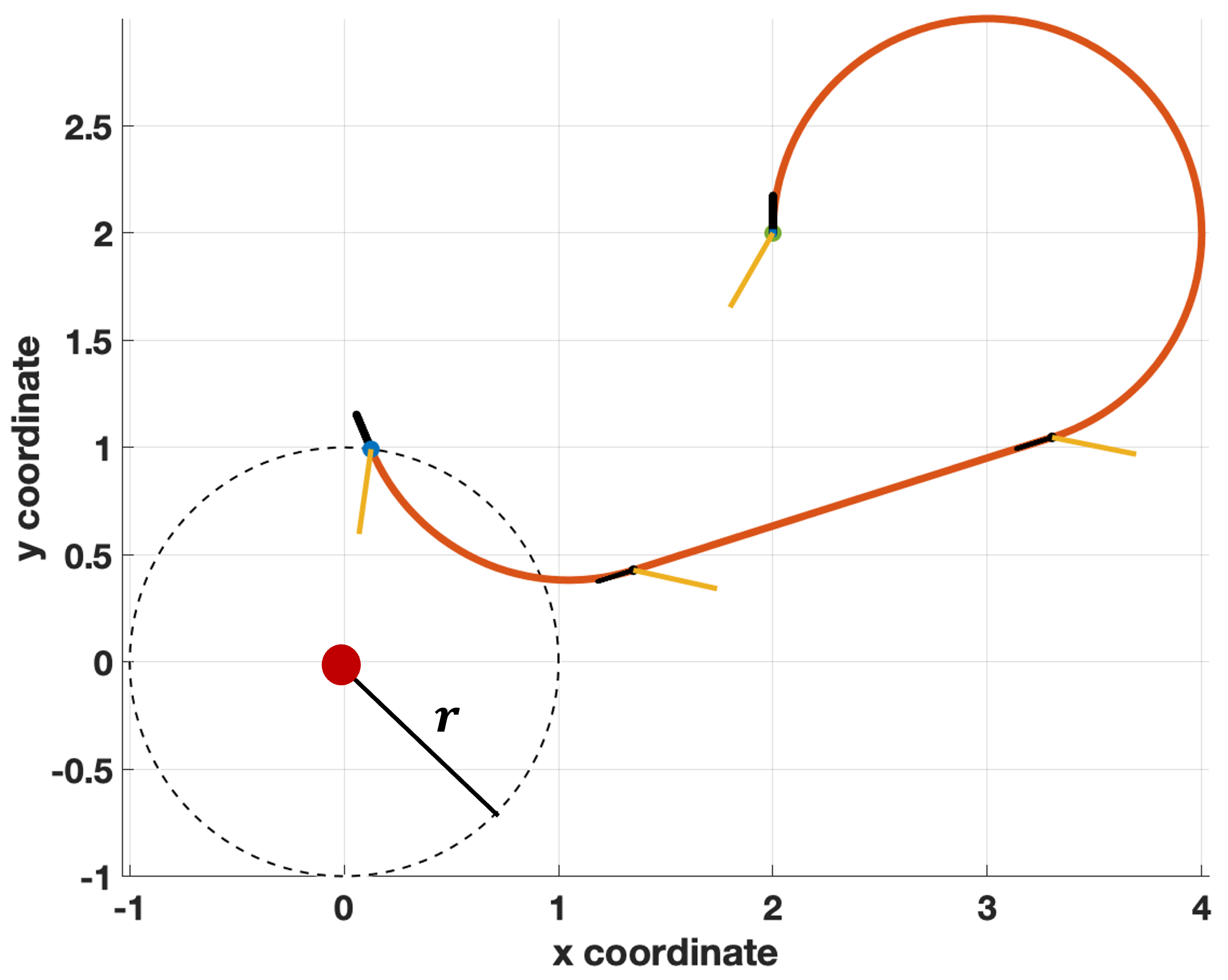}
    \caption{A $CSC$ type trajectory when $t_l^*=0$ with $x_0,y_0,\theta_0,\psi_0$, and $\omega_M$ as $2,2,\frac{\pi}{2},\frac{4\pi}{3}$, and $0.01$ respectively. The black dashed circle denotes $\partial\mathcal{C}$. The initial location is denoted as a green dot and the final location is denoted as the blue dot. The yellow line denotes the laser. }
    \label{fig:CSC_num}
\end{figure}

Finally, Figure \ref{fig:CSC_num} shows a $RSR$ trajectory when $t_l^*=0$ with $x_0,y_0,\theta_0,\psi_0$, and $\omega_M$ as $2,2,\frac{\pi}{2},\frac{4\pi}{3}$, and $0.01$, respectively. It is evident that the condition $x^2(t_f)+y^2(t_f)=r^2$ holds and the switching point and the straight line segment are collinear with the target location.

\section{Conclusion and Future Works}\label{sec:conclusion}
We considered a novel joint motion planning problem of a Dubins-Laser (Dub-L) system, in the plane, which consists of a Dubins vehicle with an attached laser. The vehicle moves with a constant unit speed and the laser, modelled as a single integrator and having a finite range, can rotate either clockwise or anti-clockwise. The aim of the Dub-L system is to capture a static target in minimum time. We characterized multiple properties of the time-optimal trajectory and established that the time-optimal trajectory is either of type $CSC$ or $CC$ or a subsegment of these two. Table \ref{tab:traj} summarizes the possible candidates of the optimal trajectory. Finally, we provide a semi-analytical solution of the shortest path which requires solving a set of at most four nonlinear equations.

Key future directions include multiple static targets in which case the problem is to determine a shortest path for the Dub-L system that captures the targets in minimum time and capturing multiple moving targets via multiple cooperative Dub-L vehicles.

\appendix
In this section, we present the proofs of the results characterized in subsection \ref{sec:charac_shortest_path}. One notational remark; we denote the time which the Dub-L system is at the first (resp. second) switching point as $t_1$ (resp. $t_2$).
\subsection{Proof of Lemma \ref{lem:no_CC}}
In this subsection, we present the proof for Lemma \ref{lem:no_CC}. We will characterize a property of a $CC$ type pose trajectory which we be instrumental in establishing the proof.

\begin{lemma}\label{lem:length_CC}
    Consider a $CC$ type pose trajectory for the Dub-L system and let $p_{\psi}=0$. Then, the length of the second $C$ segment must be greater than $\pi \rho$.
\end{lemma}
\begin{proof}
    Suppose that the length of the final $C$ segment of a $CC$ type pose trajectory of the Dub-L system is less than or equal to $\pi\rho$.
    Since $p_{\psi}=0$, from Theorem \ref{thm:costates_0_switching} and equation \eqref{eq:p_theta_trans}, we obtain $p_{\theta}(t_1)=p_{\theta}(t_f)=0$. 
    Using equation \eqref{eq:H_zero} at time $t=t_1$ and $t=t_f$, we obtain  
    \begin{equation}\label{eq:Hamil_inflex}
        \cos(\theta(t_1)-\phi)=\cos(\theta(t_f)-\phi)=\frac{-1}{\sqrt{c_x^2+c_y^2}}.
    \end{equation}
    Since $p_{\theta}(t)$ is continuous for $t\in \left(t_1,t_f\right]$ and, by equation \eqref{eq:u_opt_costates}, has a constant sign, it follows that $p_{\theta}(t)$ reaches an extremum at some point $t_3\in(t_1,t_f)$. Hence, $$\dot{p}_{\theta}(t_3)=0\implies\cos(\theta(t_3)-\phi)=\pm 1.$$ From the assumption that the length of the $C$ segment is at most $\pi \rho$, it follows that $\cos(\theta(t_3)-\phi)=-1$. From equation \eqref{eq:H_zero}, since $H(t_3)=0$, we obtain
    \begin{equation}\label{eq:Hamil_t3}
    1-\sqrt{c_x^2+c_y^2}+p_{\theta}(t_3)\frac{u(t_3)}{\rho}=0.
    \end{equation}
    Since $p_{\theta}(t_3)\frac{u(t_3)}{\rho}\leq 0$, it follows from equation \eqref{eq:Hamil_t3} that $$1\leq \frac{1}{\sqrt{c_x^2+c_y^2}}.$$ From equation \eqref{eq:Hamil_inflex}, since $\tfrac{1}{\sqrt{c_x^2+c_y^2}}$ is the cosine of an angle, this inequality can hold if $1=\sqrt{c_x^2+c_y^2}$. Substituting $1=\sqrt{c_x^2+c_y^2}$ in equation \eqref{eq:Hamil_t3} and since $u(t_3)\neq 0$ yields $p_{\theta}(t_3)=0$. Since $p_{\psi}=0$, $p_{\theta}(t_3)+p_{\psi}=0$. This means that the final segment must be a straight line segment which is a contradiction. This concludes the proof.
\end{proof}
\medskip
We now present the proof of Lemma \ref{lem:no_CC}.
\begin{figure}[t]
    \centering
\includegraphics[scale=1.2]{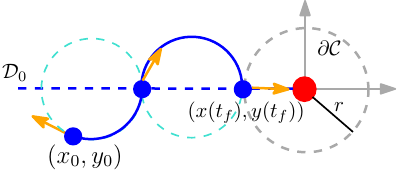}
    \caption{Illustration for the proof of Lemma \ref{lem:no_CC}. The gray dashed circle denotes $\partial\mathcal{C}$. The blue dashed line denotes the line $\mathcal{D}_0$ that passes through the switching points and the target. The blue dot denotes the Dubins vehicle of the Dub-L system and the yellow arrow denotes the laser.}
    \label{fig:no_CC}
\end{figure}
\medskip

\begin{proof}[For Lemma \ref{lem:no_CC}]
    From Theorem \ref{thm:costates_0_switching} and Lemma \ref{lem:final_loc_D0}, it follows that the circle subtended by the second $C$ segment of the $CC$ type pose trajectory must be tangent to the circle $\partial\mathcal{C}$. Note that the point of tangency is the point of intersection of line $\mathcal{D}_0$ and $\partial\mathcal{C}$ (cf. Fig \ref{fig:no_CC}).
    From the fact that the tangent point of two circles lies on the line joining the centers of the two circles, it follows that the center of the circle subtended by the second $C$ segment must also lie on the line $\mathcal{D}_0$. Next, from Theorem \ref{thm:costates_0_switching}, the switching point also lies on $\mathcal{D}_0$ which implies that the angle subtended by the second $C$ segment must be exactly $\pi$. From  Lemma \ref{lem:length_CC}, the length of the second $C$ segment must be strictly greater than $\pi\rho$. This is a contradiction and thus, a $CC$ type pose trajectory is not optimal.
\end{proof}
\subsection{Proof of Theorem \ref{thm:no_CCC}}
\begin{figure}[t]
    \centering
    \includegraphics[scale=1.15]{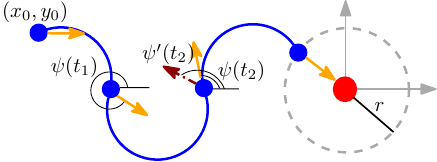}
    \caption{Illustration for proof of Lemma \ref{lem:CCC_pi_L}. The blue dot denotes the location of the Dub-L system and the yellow solid (resp. brown dashed) arrow represents the laser in trajectory $\mathcal{T}(t)$ (resp. $\mathcal{T}'(t)$). The laser turns anti-clockwise in $\mathcal{T}'(t)$ in the middle $C$ segment.}
    \label{fig:Lemma8}
\end{figure}
In this subsection, we will establish that a pose trajectory of type $CCC$ for the Dub-L system is not optimal. 

The idea is to prove the result via contradiction. Let $\mathcal{T}(t):=\Gamma(t)|\mathcal{L}(t)$ denote an optimal trajectory for the Dub-L system such that the pose trajectory $\Gamma(t)$ is of type $CCC$. In what follows, given our assumption that a trajectory $\mathcal{T}(t)$ which consists of a $CCC$ type pose trajectory is optimal, we will characterize two properties for a $CCC$ type pose trajectory. We will then use these properties to arrive at a contradiction by showing that there exists a trajectory $\mathcal{T}'(t)$ with a pose trajectory of shorter length than that of trajectory $\mathcal{T}(t)$.

\begin{lemma}\label{lem:CCC_pi_L}
    For a $CCC$ type pose trajectory, $0<\Delta\psi_2<\pi$, where 
    \begin{align*}
        \Delta \psi_2\coloneqq 
        \begin{cases}
            \psi(t_1)-\psi(t_2), \text{ if } \omega^* = \omega_M,\\
            \psi(t_2)-\psi(t_1), \text{ otherwise.}
        \end{cases}
    \end{align*}
\end{lemma}
\begin{proof}
    In this proof, we will first establish that $\Delta\psi_2<\pi$ followed by $0<\Delta\psi_2$.
    
    Let the pose trajectory $\Gamma(t)$ of trajectory $\mathcal{T}(t)$ be such that $\Delta \psi_2\geq \pi$. 
    Without loss of generality, we assume that the laser's trajectory $\mathcal{L}(t)$ is such that it turns clockwise in trajectory $\mathcal{T}(t)$. Further, when $\omega_M\leq \frac{1}{\rho}$, $\Delta\psi_2\leq 0$. Thus, given the assumption that $\Delta\psi_2\geq \pi$, $\omega_M>\frac{1}{\rho}$.
    Then, the time taken by the laser to rotate in the middle $C$ segment in trajectory $\mathcal{T}(t)$  is $\frac{\Delta\psi_2}{\omega_M-\tfrac{1}{\rho}}$. 

    Consider another trajectory $\mathcal{T}'(t):=\Gamma(t)|\mathcal{L}'(t)$ with the same
    pose trajectory $\Gamma(t)$ of type $CCC$ and the laser's trajectory $\mathcal{L}'(t)$ such that
    the laser turns anticlockwise in the middle $C$ segment and clockwise in the first and the last $C$ segment in trajectory $\mathcal{T}'(t)$ (cf. Fig. \ref{fig:Lemma8}). Mathematically, in $\mathcal{T}'(t)$, the control $\omega_{\mathcal{T}'}(t)$ of the laser is
    \begin{align*}
        \omega_{\Gamma'}(t) = 
        \begin{cases}
            \omega_M, \forall t\in [0,t_1],\\
            -\omega_M, \forall t\in (t_1,t_2],\\
            \omega_M, \forall t\in (t_2,t_f].
        \end{cases}
    \end{align*}
    Let $\Delta\psi_2'$ denote the the angle that the laser rotates in the middle $C$ segment in trajectory $\mathcal{T}'(t)$. As the pose trajectories in both $\mathcal{T}(t)$ and $\mathcal{T}'(t)$ is same,
    the total amount of time the laser rotates in the second $C$ segment must be the same in both $\mathcal{T}(t)$ and $\mathcal{T}'(t)$. This yields,
    \begin{align*}
        &\Delta \psi_2'=\frac{\Delta\psi_2(\omega_M+\tfrac{1}{\rho})}{\omega_M-\tfrac{1}{\rho}} \implies \Delta \psi_2'>\Delta\psi_2\\
        &\implies \psi'(t_2)-\psi(t_2)>0,
    \end{align*}
    where $\psi(t)$ (resp. $\psi'(t)$) denotes the laser's orientation at time $t$ in trajectory $\mathcal{T}(t)$ (resp. $\mathcal{T}'(t)$) and we used the fact that the laser is turning clockwise (resp. anti-clockwise) in trajectory $\mathcal{T}(t)$ (resp. $\mathcal{T}'(t)$). 
    
    As $\Delta\psi_2\geq \pi$ and since $\Delta\psi_2'>\Delta\psi_2$, it follows that there exists a time $t\in (t_1,t_2)$ at which $\psi'(t)=\psi(t_2)$.  This implies that $\psi'(t_2)=\psi(t_2)$ and consequently $\psi'(t_f)=\psi(t_f)$ can be achieved in trajectory $\mathcal{T}'(t)$ by having $\omega(t)=0$ for some non-zero interval of time. This implies that $t_l^*>0$ for $\mathcal{T}'(t)$. 
    This further implies, from Lemma \ref{lem:t_l>0}, that $p_{\psi}=0$. 
    This is a contradiction as, from Theorem \ref{thm:no_CC}, an optimal pose trajectory cannot be of type $CCC$ if $p_{\psi}=0$. Since pose trajectory is same in both $\mathcal{T}(t)$ and $\mathcal{T}'(t)$, trajectory $\mathcal{T}(t)$ is not optimal.

    We now show that $\Delta\psi_2>0$. Since $t_l^*=0$, $\Delta\psi_2\leq 0$ only when $\omega_M\leq \tfrac{1}{\rho}$. Given the assumption that the laser turns clockwise, it follows that $\psi(t_2)\geq \psi(t_1)$, with equality holding only when $\omega_M=\tfrac{1}{\rho}$.
    Now consider the same $\mathcal{T}'(t)$ as described earlier in this proof. Since the laser turns anti-clockwise in the middle $C$ segment in $\mathcal{T}'(t)$, it follows that by having $\omega(t)=0$ for some interval of time, $\psi'(t_f)=\psi(t_f)$ can be achieved in trajectory $\mathcal{T}'(t)$. Following analogous steps as before, it follows that there exists a trajectory of type $CS$ (or its subsegment) that requires less time than trajectory $\Gamma(t)$. This concludes the proof.
\end{proof}
\medskip
\begin{corollary}\label{cor:no_CCC}
    For any $\omega_M\leq \frac{1}{\rho}$, a pose trajectory of type $CCC$ is not optimal.
\end{corollary}
\begin{proof}
    The result follows from Lemma \ref{lem:CCC_pi_L} as $\Delta\psi_2\leq 0$ when $\omega_M\leq \frac{1}{\rho}$. 
\end{proof}
\medskip
\begin{lemma}\label{lem:CCC_pi_V}
    For a $CCC$ type pose trajectory, $0<\beta<\pi$, where $\beta$ denotes the angle subtended by the middle $C$ segment of the $CCC$ type pose trajectory.
\end{lemma}
\begin{proof}
    As a consequence of Corollary \ref{cor:no_CCC}, we only consider that $\omega_M>\frac{1}{\rho}$. 
    Let the pose trajectory $\Gamma(t)$ of trajectory $\mathcal{T}(t):=\Gamma(t)|\mathcal{L}(t)$ be of type $CCC$ for the Dub-L system and suppose that $\beta\geq \pi$. Without loss of generality, we assume that $\mathcal{L}(t)$ is such that the laser is turning clockwise in $\mathcal{T}(t)$.  
    
    Analogous to the proof of Lemma \ref{lem:CCC_pi_L}, consider the same trajectory $\mathcal{T}'(t):=\Gamma(t)|\mathcal{L}'(t)$ as described in the proof of Lemma \ref{lem:CCC_pi_L}. Note that since the pose trajectory is same in both $\mathcal{T}(t)$ and $\mathcal{T}'(t)$, the angle subtend by the middle $C$ segment in both trajectories is also equal.
    Further, since the time taken by the Dubins vehicle of the Dub-L system and the laser in the middle $C$ segment is the same, we obtain that for $\mathcal{T}'(t)$
    \begin{align*}
        \rho\beta = \frac{\Delta\psi_2'}{\omega_M+\tfrac{1}{\rho}}\implies \Delta\psi_2'=(\rho\omega_M+1)\beta,
    \end{align*}
    where $\Delta\psi_2'$ denote the the angle that the laser rotates in the middle $C$ segment in trajectory $\mathcal{T}'(t)$. As $\beta\geq \pi$ and that $\omega_M> \tfrac{1}{\rho}$, it follows that $\Delta\psi_2' \geq 2\pi$. This means that $\psi'(t_f)=\psi(t_f)$ can be achieved in trajectory $\mathcal{T}'(t)$ by having $\omega(t)=0$ for some non-zero interval of time implying that $t_l^*>0$ for $\mathcal{T}'(t)$. This further implies, from Lemma \ref{lem:t_l>0}, that $p_{\psi}=0$. Thus, analogous to the proof of Lemma \ref{lem:CCC_pi_L}, this is a contradiction. This concludes the proof. 
\end{proof}
\begin{figure}[t]
    \centering
    \includegraphics[scale=1.2]{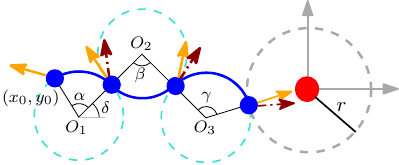}
    \caption{Illustration for the proof of Theorem \ref{thm:no_CCC}. Blue dot denotes the location of the Dub-L system for trajectory $\Gamma(t)$. Since endpoints in $\Gamma'(t)$ are same and $\epsilon_1$ and $\epsilon_2$ are infinitesimally small, the blue dot is (approximately) also the location of the Dub-L system for trajectory $\Gamma'(t)$. The yellow (resp. brown) solid (resp. dashed) arrow denotes the orientation of the laser in trajectory $\Gamma(t)$ (resp. $\Gamma'(t)$). As $\psi'(t_f')<\psi(t_f)$, $\psi'(t_f')=\psi(t_f)$ can be achieved by switching off the laser for some non-zero interval of time.}
    \label{fig:Lemma10}
\end{figure}

\medskip
We now present the proof of Theorem \ref{thm:no_CCC}.

\medskip
\begin{proof}[For Theorem \ref{thm:no_CCC}.]
    Let $\Gamma(t)$ denote a $CCC$ type pose trajectory for the Dub-L system with angles of the three $C$ segments as $\alpha,\beta$, and $\gamma$, respectively, and without loss of generality, suppose that the laser turns clockwise. Then, the time taken by the Dub-L system in $\Gamma(t)$ is 
    \begin{align*}
        T_{\Gamma}=\rho(\alpha+\beta+\gamma)=\frac{\Delta \psi_1}{\omega_M+\tfrac{1}{\rho}}+\frac{\Delta\psi_2}{\omega_M-\tfrac{1}{\rho}}+\frac{\Delta\psi_3}{\omega_M+\tfrac{1}{\rho}},
    \end{align*}
    where $\Delta \psi_1$, $\Delta \psi_2$, and $\Delta \psi_3$ denotes the angle that the laser rotates in the first, second, and third $C$ segment of $\Gamma(t)$.
    Let $\epsilon_1>0$ and $\epsilon_2>0$ be infinitesimally small positive real numbers.  Then, we can deform the pose trajectory $\Gamma(t)$ into a similar one such that the initial location, i.e., $(x(0),y(0))$,  and the final location, i.e., $(x(t_f),y(t_f))$, of the Dub-L system is same. Specifically, we can deform $\Gamma(t)$ into a similar one with the same endpoints and angles $\alpha'=\alpha+\epsilon_1$, $\beta'=\beta-\epsilon_2$, and $\gamma$. We first establish that such a deformation is possible. 
    
    Let $\boldsymbol{\nu}_{\kappa}\coloneqq \begin{bmatrix}
        \cos(\kappa) & \sin(\kappa)
    \end{bmatrix}^{\top}$ denote the unit vector of some polar angle $\kappa$ and let $\Gamma'(t)$ denote the deformed pose trajectory. Then, using Lemma \ref{lem:CCC_pi_V}, the center $O_3$ of the third $C$ segment can be expressed as
    \begin{align*}
        O_3 = O_1+2\rho\boldsymbol{\nu}_{\delta}+2\rho\boldsymbol{\nu}_{\delta-\pi+\beta},
    \end{align*}
    where $O_1$ is the center of the first $C$ segment and $\delta$ is as shown in Figure \ref{fig:Lemma10}.
    By equating the center of the third $C$ segment of trajectory $\Gamma(t)$ and $\Gamma'(t)$, we obtain
    \begin{equation}\label{eq:cos_sin_1}
    \begin{split}
        &\cos(\delta)+\cos(\delta+\pi-\beta)=\cos(\delta')+
        \cos(\delta'-\pi+\beta') \text{ and }\\
        &\sin(\delta)+\sin(\delta+\pi-\beta)=\sin(\delta')+\sin(\delta'-\pi+\beta').
        \end{split}
    \end{equation}
    Note that since $\alpha'=\alpha+\epsilon_1$, it follows that $\delta'= \delta-\epsilon_1$.
    The determinant of the Jacobian matrix obtained from the set of equations in \eqref{eq:cos_sin_1} with respect to $\epsilon_1$ and $\epsilon_2$ is 
    $-\sin(\beta-\epsilon_2)$.
    From Lemma \ref{lem:CCC_pi_V}, it follows that the determinant of the Jacobian is not zero near $\beta$ for small $\epsilon_2$. Hence, by the inverse function theorem there exists an inverse map $\mathcal{F}^{-1}:\mathbb{R}^2\to\mathbb{R}^2$ of equations \eqref{eq:cos_sin_1} such that $\mathcal{F}^{-1}$ is continuous \cite{apostol1958mathematical}. Thus, as $\epsilon_1\to 0$ and $\epsilon_2\to 0$, $\alpha'\to \alpha$ and $\beta'\to\beta$, respectively. This means that such a deformation is possible.

    Now consider one such deformation with $\epsilon_1>\epsilon_2>0$. For ease of exposition, consider a trajectory as shown in Figure \ref{fig:Lemma10}. Note that Figure \ref{fig:Lemma10} is used just for illustration and the proof can be easily modified for any such $CCC$ type trajectory. 

Let $t_1$ and $t_2$ (resp. $t_1'$ and $t_2'$) denote the time at which the Dub-L system reaches the first and the second switching point, respectively, in trajectory $\Gamma(t)$ (resp. $\Gamma'(t)$). 
In trajectory $\Gamma'(t)$, the time taken by the Dubins vehicle of the Dub-L system in the second $C$ segment is $\rho(\beta-\epsilon_2)=\tfrac{\Delta\psi'_2}{\omega_M-\tfrac{1}{\rho}}$. This implies that $\Delta\psi_2' = \Delta\psi_2-\rho\epsilon_2(\omega_M-\frac{1}{\rho})$, where we used the fact that $\rho\beta=\frac{\Delta\psi_2}{\omega_M-\tfrac{1}{\rho}}$. 
As the laser turns clockwise, we obtain 
    \begin{align*}
        & \psi'(t_1')-\psi'(t_2') = \psi(t_1)-\psi(t_2)-\rho\epsilon_2\left(\omega_M-\frac{1}{\rho}\right),\\
        \implies& \psi'(t_2')-\psi(t_2) = \psi'(t_1')-\psi(t_1)+\rho\epsilon_2\left(\omega_M-\frac{1}{\rho}\right),\\
        \implies& \psi'(t_2')-\psi(t_2) =-\rho\omega_M(\epsilon_1-\epsilon_2)-\epsilon_1-\epsilon_2<0,
    \end{align*}
    where we used that $\Delta\psi_1' = \psi(0)-\psi'(t_1')$ and $\Delta\psi_1 = \psi(0)-\psi(t_1)$ in the second step.

    If the pose trajectory $\Gamma(t)$ is optimal, then $T_{\Gamma}<T_{\Gamma'}$. This implies
    \begin{align*}
        & \frac{\Delta\psi_1-\Delta\psi_1'}{\omega_M+1/\rho}+\frac{\Delta\psi_2-\Delta\psi'_2}{\omega_M-1/\rho}+\frac{\Delta\psi_3-\Delta\psi'_3}{\omega_M+1/\rho}<0\\
        &\implies \rho(\omega_M+\tfrac{1}{\rho})(\epsilon_2-\epsilon_1)+\Delta\psi_3-\Delta\psi'_3<0\\
        &\implies \rho(\omega_M+\tfrac{1}{\rho})(\epsilon_2-\epsilon_1) + \psi(t_2)-\psi'(t_2')+\\
        &\qquad\psi'(t_f')-\psi(t_f)<0\\
        &\implies \psi'(t_f')-\psi(t_f)<-2\epsilon_2<0.
    \end{align*}
    This is a contradiction because, as $\psi'(t_f')-\psi(t_f)<0$, $\psi'(t_f')=\psi(t_f)$ can be achieved by having $\omega(t)=0$ for some non-zero interval of time implying that $t_l^*>0$ for $\Gamma'(t)$. This further implies, from Lemma \ref{lem:t_l>0}, that $p_{\psi}=0$ for $\Gamma'(t)$.
    From Theorem \ref{thm:no_CC}, this implies that there exists a $CS$, $C$ or an $S$ type trajectory from the same initial location which requires less time than pose trajectory $\Gamma'(t)$ (and consequently $\Gamma(t)$). Thus, $\Gamma(t)$ is not optimal.
\end{proof}

\subsection{Proof of Theorem \ref{thm:CSC_on_C}}

\begin{figure}[t]
    \centering
    \includegraphics[scale=1.3]{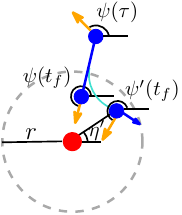}
    \caption{Illustration for Case 1 of proof of Theorem \ref{thm:CSC_on_C}. The blue line denotes the pose trajectory $\Gamma(t)$ and the cyan curve denotes the pose trajectory $\Gamma'(t)$.}
   \label{fig:S_on_circle}
\end{figure}

\begin{figure}[t]
    \centering
    \includegraphics[scale=1.4]{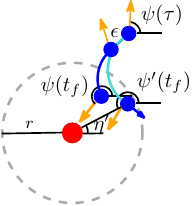}
    \caption{Illustration for Case 2 of proof of Theorem \ref{thm:CSC_on_C}. The blue line denotes the pose trajectory $\Gamma(t)$ and the cyan curve denotes the pose trajectory $\Gamma'(t)$.}
   \label{fig:C_on_circle}
\end{figure}
\begin{proof}
    Since the trajectory of the Dub-L system may end with an arc or a straight line, we consider two cases based on whether the trajectory ends with an $S$ or a $C$. For both cases, we assume that the laser is turning anti-clockwise. The proof when the laser turns clockwise is analogous.

    \textbf{Case 1 (Trajectory is of $S$ type at $t_f$):} Consider a trajectory $\mathcal{T}(t):=\Gamma(t)|\mathcal{L}(t)$, as shown in Figure \ref{fig:S_on_circle}, with the pose trajectory $\Gamma(t)$ such that it ends with an $S$ segment and for which $x^2(t_f)+y^2(t_f)<r^2$ holds. 
    Consider another trajectory $\mathcal{T}'(t):=\Gamma'(t)|\mathcal{L}(t)$. The pose trajectory $\Gamma'(t)$ 
    has the same initial condition as $\Gamma(t)$ and, because of Lemma \ref{lem:same_sign}, has the vehicle turn anti-clockwise at time $\tau=t_f-\epsilon$, where $\epsilon>0$ is a very small positive real number. For all $t<\tau$, $\Gamma'(t)$ is the same as $\Gamma(t)$.
    Further, $\Gamma'(t)$ is such that the time taken by the Dub-L system in both $\mathcal{T}(t)$ and $\mathcal{T}'(t)$ is equal. 
    Since both $\Gamma(t)$ and $\Gamma'(t)$ are same until time $\tau$ and since the laser's trajectory is same in both $\mathcal{T}(t)$ and $\mathcal{T}'(t)$, it follows that $\psi(\tau)=\psi'(\tau)$, where $\psi(t)$ (resp. $\psi'(t)$) denotes the orientation of the laser at time $t$ in trajectory $\mathcal{T}(t)$ (resp. $\mathcal{T}'(t)$). 
    Note that it is possible that the constraint specified in equation \eqref{eq:terminal_constraints_b} is not satisfied in $\mathcal{T}'(t)$ since we keep the time taken by both trajectories equal.
    As the time taken by both trajectories is  equal, we obtain
    \begin{align}
        &\frac{\psi'(t_f)-\psi(\tau)}{\omega_M+\tfrac{1}{\rho}} = \frac{\psi(t_f)-\psi(\tau)}{\omega_M}\nonumber\\
        &\Rightarrow \psi'(t_f)-\psi(t_f)=\frac{\psi(t_f)-\psi(\tau)}{\omega_M\rho}\geq 0, \label{eq:S_onC}
    \end{align}
    where we used the fact that the laser is turning anti-clockwise.

    Let $\eta$ (resp. $\eta'$) denote the angle from the target to the final location of the Dub-L system in trajectory $\mathcal{T}(t)$ (resp. $\mathcal{T}'(t)$). Then, from construction of $\Gamma'(t)$, we have that $\eta'-\eta<0$ (cf. Fig. \ref{fig:S_on_circle}).
    Further, if trajectory $\mathcal{T}(t)$ is optimal, since the vehicle turns anti-clockwise in $\Gamma'(t)$, $\psi'(t_f)-\pi\leq \eta'$ must hold. This is because if $\psi'(t_f)-\pi>\eta'$, then by having $\omega(t)=0$ for some non-zero interval of time, $\psi'(t_f)-\pi=\eta'$ can be achieved implying that $t_l^*>0$ for $\Gamma(t)$. This further implies, from Lemma \ref{lem:on_C} and Lemma \ref{lem:t_l>0}, that there exists an optimal trajectory that ends on $\partial \mathcal{C}$ and hence, $\Gamma(t)$ is not optimal. Thus, using that $\psi'(t_f)-\pi\leq \eta'$, equation \eqref{eq:S_onC} yields
    \begin{align*}
        \eta'+\pi-\psi(t_f)\geq 0\implies
        \eta'-\eta\geq 0,
    \end{align*}
    which is a contradiction.

    \textbf{Case 2 (Trajectory is of $C$ type at $t_f$):}
    Consider a trajectory $\mathcal{T}:=\Gamma(t)|\mathcal{L}(t)$ with a pose trajectory $\Gamma(t)$ (Fig. \ref{fig:C_on_circle}) which ends with a circular arc at time instant $t_f$ and for which $x^2(t_f)+y^2(t_f)<r^2$ holds. 
    Consider another trajectory $\mathcal{T}'(t):=\Gamma'(t)|\mathcal{L}(t)$ with pose trajectory $\Gamma'(t)$.
    The pose trajectory $\Gamma'(t)$ is such that it has the vehicle turn anti-clockwise at time $\tau=t_2-\epsilon$, where $t_2$ denotes the time instant when the vehicle switches from an $S$ segment to the final $C$ segment in $\Gamma(t)$. For any time $t<\tau$, $\Gamma'(t)$ is the same as $\Gamma(t)$. Further, $\Gamma'(t)$ is such that the time taken by the Dub-L system in $\Gamma'(t)$ is equal to the time taken by the Dub-L system in $\Gamma(t)$.
    As the time taken by the vehicle is equal in both trajectories, we obtain
    \begin{align}
        &\frac{\psi'(t_f)-\psi(\tau)}{\omega_M+\tfrac{1}{\rho}} = \epsilon + \frac{\psi(t_f)-\psi(\tau)-\epsilon \omega_M}{\omega_M+\tfrac{1}{\rho}}\nonumber\\
        &\Rightarrow \psi'(t_f)-\psi(t_f)=\frac{\epsilon}{\rho}>0. \label{eq:C_onC}
    \end{align}
    With analogous arguments as in Case 1, it can be shown that if $\Gamma(t)$ is optimal then, $\eta'-\eta>0$ which is a contradiction. This concludes the proof.
\end{proof}
\section*{References}



\bibliographystyle{ieeetr}
\bibliography{references}
\end{document}